\documentclass{article}

\usepackage{microtype}
\usepackage{graphicx}
\usepackage{subcaption}
\usepackage{booktabs} 
\usepackage{multirow}
\usepackage{amsmath}
\usepackage{tabularx}
\usepackage{bbm}
\usepackage[bb=boondox]{mathalfa}
\usepackage{hyperref}
\usepackage{fancyhdr}

 \usepackage[accepted]{icml2025}

\usepackage{amsmath}
\usepackage{amssymb}
\usepackage{mathtools}
\usepackage{amsthm}
\usepackage{multirow}
\usepackage{bm}
\usepackage{adjustbox}
\usepackage[capitalize,noabbrev]{cleveref}

\theoremstyle{plain}

\theoremstyle{definition}

\theoremstyle{remark}

\usepackage[textsize=tiny]{todonotes}

\begin{document}
\icmltitlerunning{Interpolating Neural Network-Tensor Decomposition (INN-TD)}

\twocolumn[
\icmltitle{Interpolating Neural Network-Tensor Decomposition (INN-TD): a scalable and interpretable approach for large-scale physics-based problems}




\begin{icmlauthorlist}
\icmlauthor{Jiachen Guo}{yyy}
\icmlauthor{Xiaoyu Xie}{comp}
\icmlauthor{Chanwook Park}{comp}
\icmlauthor{Hantao Zhang}{yyy}
\icmlauthor{Matthew J. Politis}{yyy}
\icmlauthor{Gino Domel}{comp}
\icmlauthor{Thomas J.R. Hughes}{tx}
\icmlauthor{Wing Kam Liu}{comp,hidenn}
\end{icmlauthorlist}

\icmlaffiliation{yyy}{Theoretical and applied mechanics program, Northwestern University, 2145 Sheridan Rd, Evanston, Illinois, USA}
\icmlaffiliation{comp}{Department of mechanical engineering, Northwestern University, 2145 Sheridan Rd, Evanston, Illinois, USA}
\icmlaffiliation{hidenn}{HIDENN-AI, Evanston, Illinois, USA}
\icmlaffiliation{tx}{Institute for Computational Engineering and Sciences, The University of Texas at Austin, Austin, Texas, USA}

\icmlcorrespondingauthor{Wing Kam Liu}{w-liu@northwestern.edu}

\icmlkeywords{Machine Learning, ICML}

\vskip 0.3in
]



\printAffiliationsAndNotice{} 

\begin{abstract}
Deep learning has been extensively employed as a powerful function approximator for modeling physics-based problems described by partial differential equations (PDEs). Despite its popularity, standard deep learning models often demand prohibitively large computational resources and yield limited accuracy when scaling to large-scale, high-dimensional physical problems. Their black-box nature further hinders the application in industrial problems where interpretability and high precision are critical. To overcome these challenges, this paper introduces Interpolating Neural Network-Tensor Decomposition (INN-TD), a scalable and interpretable framework that has the merits of both machine learning and finite element methods for modeling large-scale physical systems.  By integrating locally supported interpolation functions from the finite element into the network architecture, INN-TD achieves a sparse learning structure with enhanced accuracy, faster training/solving speed, and reduced memory footprint. This makes it particularly effective for tackling large-scale high-dimensional parametric PDEs in training, solving, and inverse optimization tasks in physical problems where high precision is required.  
\end{abstract}

\section{Introduction}
Deep learning methods have become a popular approach to model physical problems in recent years. Once trained, these models can make predictions much faster than standard numerical simulations. However, deep learning faces challenges in modeling physical problems such as lower accuracy, instability, and the need for a huge amount of high-fidelity training data. As a result, its applications remain limited for many engineering fields that require high precision, such as aerospace, semiconductor manufacturing, and earthquake engineering. In this paper, we propose Interpolating Neural Network-Tensor Decomposition (INN-TD), which leverages the merits of both machine learning and numerical analysis to tackle large-scale, high-dimensional physical problems governed by partial differential equations (PDEs).

\section{Related work}
Deep learning theory has been applied to solve physical problems that are challenging for standard numerical algorithms. Based on the objective of the task, the current works can be broadly classified into 3 major categories: data-driven trainers, data-free solvers, and inverse problems. 

\subsection{Data-driven trainers}
In the data-driven regime, we aim to leverage the universal approximation capabilities of neural networks to approximate PDE solutions.  Multilayer perceptrons (MLP) have been used due to their universal approximation capability \cite{cybenko1989approximation}. Depending on the shape of the domain, convolutional neural networks and graph neural networks can be used as additional inductive biases for data on regular grids and irregular geometries, respectively \cite{gao2021phygeonet,pfaff2020learning}. In recent years, many operator learning methods have been proposed to learn functional mappings instead of functions \cite{li2020fourier,lu2021learning,huang2023introduction}. However, due to the black-box nature, designing optimal model architectures and optimization hyperparameters is still largely heuristic and requires trial and error \cite{10.5555/3692070.3693785,colbrook2022difficulty}. For problems where physics is fully known, it becomes cumbersome to first generate offline data using standard numerical solvers and then train surrogate models from the offline data. Data generation and storage for large-scale physical problems can be extremely expensive. For these problems, a better approach is to directly solve PDEs using deep learning models.
\subsection{Data-free solvers}
For cases where no data is available, deep learning can be used as function approximators to approximate PDE solutions. For example, MLP has been vastly used in physics-informed neural networks (PINNs) and its variations to approximate solutions to PDEs \cite{raissi2019physics, zhang2022analyses}. Despite their vast usage, however, unlike classical numerical solvers which have a solid theoretical foundation, most of the current deep learning models are based on heuristics and lack theoretical analysis \cite{wang2022and}. It has been shown that PINN results have often been compared to weak baselines, potentially overestimating their capabilities \cite{mcgreivy2024weak}. Most data-free deep learning solvers also suffer from the curse of dimensionality \cite{krishnapriyan2021characterizing} and are relatively slow compared to standard numerical solvers such as finite difference/volume/element methods \cite{grossmann2024can}.
\subsection{Finite element interpolation}
Finite element method (FEM)  is the computational workhorse to numerically solve differential equations arising in physics-based simulation \cite{liu2022eighty}. FEM is based on finite element (FE) interpolation theory. FEM first decomposes the original computational domain into a finite number of elements and then leverages locally-supported FE interpolation functions to approximate the field variables within each local domain. FEM can solve a wide range of problems, including structural mechanics, heat transfer, fluid flow, weather forecasting, and, in general, every conceivable problem that can be described by PDEs. Recent advancements in finite element (FE) methods focus on utilizing machine learning to enhance traditional FE interpolation theory. By employing Convolutional-Hierarchical Deep Neural Network (C-HiDeNN), a generalized FE interpolation framework has been proposed, offering improved flexibility and greater accuracy \cite{lu2023convolution, li2023convolution, park2023convolution}. Despite their robustness and widespread application, finite element (FE)-type methods have been primarily applied to address space/space-time problems, exhibiting limitations in handling parametric partial differential equations (PDEs). Furthermore, FE methods are particularly susceptible to the curse of dimensionality, which poses significant computational challenges and escalates costs when applied to large-scale, high-dimensional problems \cite{bungartz2004sparse}.

\subsection{Tensor decomposition}
Tensor decomposition (TD) has been widely used in data compression tasks \cite{sidiropoulos2017tensor}. Among different tensor decomposition schemes, the Canonical Polyadic (CP) decomposition has been very popular since it decomposes higher-order tensors into a tensor product of 1D vectors \cite{kolda2009tensor}. CP decomposition has also been used vastly to solve PDEs \cite{beylkin2005algorithms,bachmayr2023low,dolgov2012fast,cohen2010convergence}. From now on, we refer to CP decomposition as TD unless stated otherwise. TD has also been used in regression and classification tasks as an efficient model due to a lesser number of parameters \cite{ahmed2020tensor, park2024engineering}. Recently, it has been proved that if MLP is used to approximate each mode in TD, then it’s also a universal approximation \cite{vemuri2025functional}.

\section{INN-TD formulation}
Interpolating Neural Network-Tensor Decomposition (INN-TD) achieves its remarkable efficiency through two key components: the utilization of C-HiDeNN, which is a general locally supported interpolation theory based on finite element, and the integration of tensor decomposition techniques. These elements collectively enhance the framework's capability to model complex physical systems with improved accuracy and computational efficiency. 

\subsection{C-HiDeNN interpolation}

\begin{figure}[h]
\centering
\includegraphics[width=0.9\linewidth]{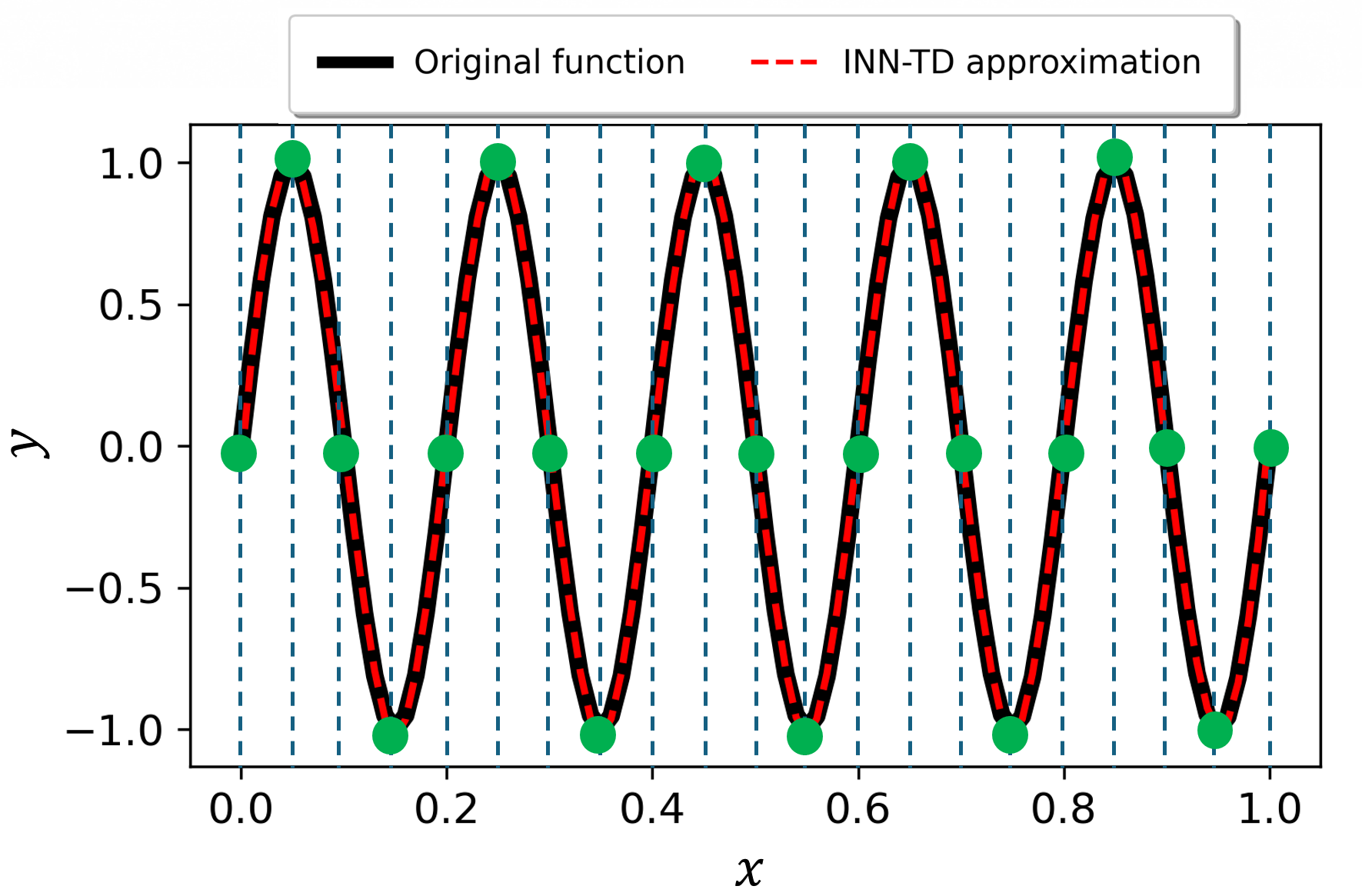}
\caption{C-HiDeNN approaximation based on locally supported basis functions}
\label{sine}
\end{figure}

Assume we want to learn a univariate function $f(x)$ given data $(x_i, y_i)$. In INN-TD, the target function $f(x)$ is first broken into small elements, and then data within each element is approximated locally. This concept has been widely used in numerical methods such as finite element (FE). Here, we use a generalized version of FEM shape functions called Convolutional-Hierarchical Deep Neural Network (C-HiDeNN) \cite{lu2023convolution} as the basis for the local interpolation function. For example, to approximate a simple sinusoidal function over multiple periods as shown in Fig. \ref{sine}, we start by dividing the original domain into several elements, which we refer to as a mesh. The boundaries of each element are referred to as nodes, which are highlighted as green dots in Fig. \ref{sine}. For each segment of the mesh, we employ local interpolation functions to perform the approximation locally. The final approximation of the entire function is achieved by linking these locally approximated segments together.

Assuming approximating the original function using $n$ elements, the univariate function $f(x)$ is approximated using C-HiDeNN interpolation functions in INN-TD, and can be written as:

\begin{equation}
    f(x) \approx \widetilde{\boldsymbol{N}}(x;s,a,p)\boldsymbol{u}_{x}
    \label{chidenn}
\end{equation}

where $\boldsymbol{u}_{x}\in \mathbb{R}^{n + 1}$ is defined as the nodal values (grid point values); $\widetilde{\boldsymbol{N}}(x;s,a,p)$ is the C-HiDeNN interpolation basis functions for different $n$ elements as shown in Eq. \ref{node}. The approximation is governed by three hyperparameters, which allow for the approximation of polynomial orders of any degree: patch size $s$, reproducing order $p$, and dilation parameter $a$.  Details regarding the role of each hyperparameter are provided in the Appendix. \ref{sec:append_chidenn}. 

\begin{equation}
    \widetilde{\boldsymbol{N}}(x;s,a,p) = [\widetilde{N}_0(x;s,a,p), ..., \widetilde{N}_n(x;s,a,p)]
    \label{node}
\end{equation}

C-HiDeNN interpolation function maintains all the essential finite element approximation properties such as  Kronecker delta \cite{hughes2003finite}, partition of unity \cite{melenk1996partition}, error estimations and convergence theorems \cite{lu2023convolution}. 
For example, $\widetilde{\boldsymbol{N}}(x)$ satisfies the following Kronecker delta property at nodal position $x_l$:
\begin{equation}
    \widetilde{N}_k(x_l;s,a,p)=\delta_{kl}
\end{equation}
where the Kronecker delta is defined as:
\begin{equation}
    \delta_{kl}=\begin{cases}0&\mathrm{if~}k\neq l,\\1&\mathrm{if~}k=l.\end{cases}
\end{equation}
At the Dirichlet boundary node ($x_l$), we have:
\begin{equation}
    \sum_{k}\widetilde{N}_k(x_l; a, s, p)u_k=u_l
\end{equation}
Thus, unlike many data-free solvers where an additional penalty term has to be added to softly enforce the Dirichlet boundary condition, C-HiDeNN can easily enforce the Dirichlet boundary condition. 

In summary, C-HiDeNN basis function has the following advantages over MLP, especially when applied to physical problems: 1. The Dirichlet boundary condition is automatically satisfied. 2. C-HiDeNN is interpretable as it uses locally supported basis functions and the number of elements in C-HiDeNN is closely related to the resolution. 3. C-HiDeNN facilitates numerical integration through the use of Gaussian quadrature \cite{hughes2003finite}, making it straightforward and efficient for many classical integral methods in solving PDEs. For simplicity of the notation, hyperparameters $s,a,p$ are dropped from now on.  

\subsection{INN-TD approximation}
A general multivariate function $u(x_1,x_2,…,x_D)$ can be approximated using functional tensor decomposition (TD) as \footnote{For ease of discussion, we only cover cases where the outputs are scalars in the main text. Nonetheless, the current framework seamlessly extends to handle high-dimensional outputs, as shown in Appendix \ref{sec:elasticity}.}
\begin{equation}
    u(x_1,x_2,...,x_D)\approx \mathcal{J} u(\bm{x}) = \sum_{m=1}^M \Pi_{d=1}^D f_d^{(m)}(x_d)
    \label{td_i}
\end{equation}

where $M$ is defined as the total number of modes in TD; $D$ is the total number of input dimensions; $\bm{x} = (x_1,x_2,…,x_D)$; $f_d^{(m)}(x_d)$ is the univariate function for $d$-th dimension and $m$-th mode. It has been proved that when the mode number $M$ is sufficiently large and each univariate function $f_d^{(m)}(x_d)$ serves as a universal approximation, Eq. \ref{td_i} becomes a universal approximation for any multivariate function \cite{vemuri2025functional}. 

In INN-TD, we utilize the C-HiDeNN interpolation function for each univariate function $f_d^{(m)}(x_d)$. As a result, Eq. \ref{td_i} can be written as:
\begin{equation}
    \mathcal{J}u(\bm{x})=\sum_{m=1}^{M}\widetilde{\boldsymbol{N}}_{1}(x_1)\boldsymbol{u}_{x_1}^{(m)}\cdot\widetilde{\boldsymbol{N}}_2(x_2)\boldsymbol{u}_{x_2}^{(m)}\cdot... \cdot\widetilde{\boldsymbol{N}}_D(x_D)\boldsymbol{u}_{x_D}^{(m)}
    \label{td_eq}
\end{equation}

The model parameters for each mode in the INN-TD framework are $\bm{U}^{(m)} = [\bm{u}_1^{(m)}, \bm{u}_2^{(m)}, ..., \bm{u}_D^{(m)}]$. Since we have $M$ modes in the INN-TD interpolation, the complete model parameters are represented as ${\bm{U} = [\bm{U}^{(1)}, \bm{U}^{(2)}, ..., \bm{U}^{(M)}]}$. The overall structure of INN-TD is shown in Fig. \ref{inn-td} in the Appendix.

\subsection{Interpretability of INN-TD}
As stated in \cite{doshi2017towards}, model interpretability refers to the ability to explain a model’s behavior in a way that is understandable to humans. INN-TD is an interpretable machine learning algorithm designed for scientific and engineering problems, offering transparency in its structure, clear convergence guarantees, and explainable insight into how data or parameter changes affect outcomes. In addressing physics-based problems described using PDEs, we harness several inductive biases inherent in both physics and classical numerical methods. Many physical problems tend to be low-dimensional in nature. For instance, in structural engineering, it is often observed that only the initial few frequencies significantly influence a structure’s dynamic response. As such, this response can be effectively approximated using a limited number of modes. More broadly, proper orthogonal decomposition (POD) or Karhunen-Loève expansions are widely employed to simplify the complexity inherent in original physical models and have been widely used in reduced-order modeling of structural analysis and fluid dynamics problems \cite{rathinam2003new}. Consequently, we utilize INN-TD as an efficient tool to uncover the naturally low-dimensional characteristics of physical problems.

Additionally, INN-TD incorporates the concept of locally supported interpolation functions from the finite element method as approximators. Specifically, we employ the C-HiDeNN interpolation function, which balances the robust capabilities of generalized finite element methods with the adaptability of machine learning approaches. As a result, INN-TD adeptly accounts for the Kronecker Delta and partition of unity properties, providing flexibility, locality, stability, and control over the interpolation process. Detailed definitions of these desired properties are explained in Appendix \ref{sec:append_chidenn}. As a result, INN-TD is very efficient and accurate in terms of capturing complex physics for both data-driven training and data-free solving tasks.

\begin{figure}[h]
\centering
\includegraphics[width=0.7\linewidth]{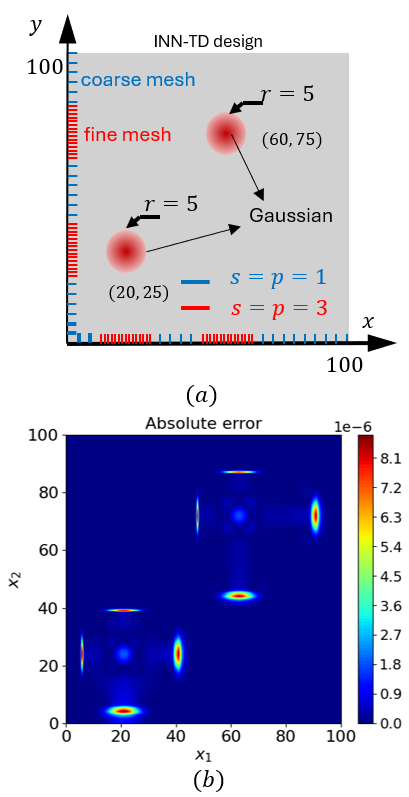}
\caption{Interpretable design of INN-TD: (a) the 1D meshes used in INN-TD are only refined at the location where nonlinearity happens with larger $s$ and $p$ used to achieve higher-order smoothness (b) point-wise absolute error between INN-TD and the exact solution}
\label{interp11}
\end{figure}

Finally, unlike many black-box deep learning methods, INN-TD offers a clear interpretation. Specifically, the number of elements $n_d$ in dimension $d$ determines the resolution of the interpolation, with larger $n_d$ corresponding to a higher-resolution interpolation in dimension $d$. For cases where $D = 2$, the total number of modes $M$ is closely linked to the number of singular values in singular value decomposition (SVD) \cite{kolda2009tensor}. The benefits of such interpretability when solving PDEs are demonstrated in detail through a numerical example in Fig. \ref{interp11}. In this example, INN-TD is leveraged to solve the 2D Poisson's equation with the local source term. By controlling mesh density and hyperparameters $s$ and $p$, INN-TD can accurately predict the solution field. Details of this example can be found in Appendix \ref{sec:Inter}. 

Moreover, due to its interpretable nature, INN-TD exhibits convergence properties as a data-free solver. Specifically, increasing the number of model parameters tends to improve accuracy. This is corroborated using multiple numerical examples in Section \ref{sec:solving}. The error bound is also theoretically shown in \cite{guo2025tensor}.

\subsection{Data-driven training}
For training tasks, our goal is to estimate the function $y \approx \mathcal{J}u(\bm{x}) $ given the data pairs $(\bm{x}_i, y_i)$. In this section, we focus solely on regression tasks, as our primary interest lies in learning function mapping to approximate physical relationships \footnote{Data-driven training code can be found in: {\scriptsize\url{https://github.com/hachanook/pyinn/tree/icml_rev}}}. The model parameters ${\bm{U}}$ can be obtained via a standard data-driven framework. We utilize the mean squared error (MSE) as the loss function to measure the difference between predicted values and true labels:
\begin{equation}
    \mathcal{L} = \frac{1}{K}\sum_k(\mathcal{J}u(\bm{x}_k^*; \bm{U})-y_k^*)^2.
    \label{train_boost}
\end{equation}
where $k$ is the index of training data; $K$ is the total number of training data; $(\bm{x^*,} y^*)$ is the training data pairs.

We propose two training schemes for determining model parameters. The first scheme utilizes a boosting algorithm, treating each mode as a weak learner. We iteratively introduce new weak learner $\mathcal{J}^{(M)} u\bm{{(x_k^*; U}}) $ until the loss meets the predefined criteria:

\begin{equation}
\begin{aligned}
    \operatorname*{argmin}_{\bm{U}^{(M)}} \mathcal{L} = &\operatorname*{argmin}_{\bm{U}^{(M)}} \frac{1}{K}\sum_k(\Sigma_{m=1}^{M-1} \mathcal{J}^{(m)} u(\bm{x}_k^*; \bm{U}^{(m)}) + \\ &\mathcal{J}^{(M)} u(\bm{x}_k^*; \bm{U}^{(M)}) -y_k^*)^2.
    \label{train_boost}
    \end{aligned}
\end{equation}

In the second scheme, known as all-at-once optimization, model parameters for all modes are determined simultaneously.

\begin{equation}
    \operatorname*{argmin}_{\bm{U}} \mathcal{L} = \operatorname*{argmin}_{\bm{U}}\frac{1}{K}\sum_k(\mathcal{J}u(\bm{x}_k^*; \bm{U})-y_k^*)^2.
    \label{train_boost}
\end{equation}
Details about the boosting and all-at-once training algorithms can be found in Appendix \ref{sec:append_trainer}.
\subsection{Data-free solving}
Contrary to data-driven training tasks, no data is available to train the model parameters for data-free solving. However, we know the underlying physics is governed by the corresponding PDE. This knowledge can be leveraged to solve the model parameters in an a priori sense so that the PDE and its initial and boundary conditions are satisfied. As a result, data-free solving tasks are significantly more challenging than data-driven training tasks. \footnote{Data-free solving code can be found in: {\scriptsize\url{https://hub.docker.com/r/chiachenkuo/spt_demo}}}

In the solving task, we categorize the independent variables $\bm{x}$ into three distinct types based on their physical significance. These are: spatial variables $\bm{x}_s$, which describe geometric features; the temporal variable $x_t$; and parametric variables $\bm{x}_p$, which may include PDE coefficients, forcing terms, as well as initial and boundary conditions. Thus, a general PDE for physical problems can be represented as a space-parameter-time (S-P-T) problem.
\begin{equation}
    \mathcal{L}(\mathcal{J}u(\bm{x}_s, \bm{x}_p, {x}_t))=0,
\end{equation}

where $\mathcal{L}(\cdot)$ denotes a general (nonlinear) differential operator; $\mathcal{J}u(\bm{x}_s, \bm{x}_p, {x}_t)$ is the approximated solution, commonly referred to as the trial function in numerical analysis.

Methods for solving partial differential equations (PDEs) generally fall into two categories. The first involves directly minimizing the residual of the PDE in its differential form. The second projects the PDE loss onto a subspace, evaluating the loss via a weighted-sum residual in the integral form. Similar to FEM, INN-TD employs the integral form, which benefits from the efficiency of Gaussian quadrature-based numerical integration:

\begin{equation}
    \int\delta\mathcal{J}u(\bm{x}_s, \bm{x}_p, {x}_t)\mathcal{L}(\mathcal{J}u(\bm{x}_s, \bm{x}_p, {x}_t))d\bm{x}_s d\bm{x}_p d{x}_t=0
    \label{galerkin0}
\end{equation}

where $\delta \mathcal{J}u(\bm{x}_s, \bm{x}_p, {x}_t)$ is called as the test function \cite{hughes2003finite}. Different test functions can be adopted depending on the formulation. In this paper, we use the Galerkin formulation where the same function space is used for trial and test functions \cite{hughes2003finite}. Additionally, we leverage integration by parts in Eq. \ref{galerkin0} to relax the requirement for function continuity in INN-TD approximation \cite{guo2024convolutional}.

In solving high-dimensional S-P-T partial differential equations (PDEs), a major challenge is the exponential increase in computational cost as the number of dimensions rises. This phenomenon, known as the curse of dimensionality, has been one of the most critical bottlenecks for both standard numerical and deep learning-based solvers. To address this challenge, only one-dimensional integration is performed in Eq. \ref{galerkin0} due to the tensor decomposition structure of the INN-TD approximation. This approach ensures that the number of unknowns in the resulting system of equations depends only on the grid points across each dimension. Consequently, INN-TD effectively circumvents the curse of dimensionality, allowing it to handle high-dimensional problems using very fine grids and achieving high resolution. This capability is especially advantageous for modeling physical phenomena that demand an extremely fine mesh to accurately capture and resolve localized features.

In this paper, Eq. \ref{galerkin0} is solved using a boosting algorithm. Moreover, instead of relying on stochastic optimization schemes, we use subspace iteration to linearize Eq. \ref{galerkin0}, taking advantage of the sparsity in the resulting system of equations, and solve it using sparse direct solvers. Details about the data-free solver algorithm can be found in Appendix \ref{sec:append_solver}.

\subsection{Inverse optimization}
In many engineering applications, determining the optimal coefficients or design parameters from experimental data or underlying physics is essential. In such cases, INN-TD serves as an efficient parametric mapping tool from the input space to the output space, thanks to its sparse structure.

There are, in general, two different ways to tackle this problem. In the first case, the spatial, temporal, and parametric variables are all treated as inputs to the machine learning model. Once the model is trained/solved, one can easily leverage automatic differentiation to infer the parametric inputs \cite{takamoto2022pdebench}. This approach requires the model to accurately learn the parametric mapping. In the second case, the physical parameters become trainable neural network parameters. While they don't appear directly in the network, they influence the loss function through an additional residual \cite{zhang2022analyses}. The performance of this approach can be sensitive to the contribution of each loss term and may require adaptive scaling of the loss function components \cite{berardi2025inverse}.

In this paper, we adopt the first approach for INN-TD to solve inverse problems, as INN-TD can efficiently and accurately handle general parametric function mapping. The optimal parameters can be obtained by minimizing the following loss function.

\begin{equation}
    \operatorname*{argmin}_{\bm{x_p}} \mathcal{L} =\operatorname*{argmin}_{\bm{x_p}}\|\mathcal{J}u(\bm{x_s}, x_t; \bm{x_p}) - u^{*}(\bm{x_s}, x_t)\|_{\ell_2}
\label{inverse_loss}
\end{equation}
where $\|\cdot\|_{\ell_2}$ is the L2 norm and is defined as $\| \bm{x} \|_{\ell_2} = \sqrt{\sum_{i=1}^n |x_i|^2}$; $u^{*}(\bm{x_s}, x_t)$ is the target spatial-temporal field.

\section{Results}

In this section, we present the performance of INN-TD in three key areas: data-driven training, data-free solving, and inverse optimization. We compare its performance with other deep learning models including vanilla MLP, SIREN \cite{sitzmann2020implicit}, Kolmogorov-Arnold networks (KAN) \cite{liu2024kan}, and tensor-decomposition physics-informed neural networks (CP-PINN) \cite{vemuri2025functional}. All of the cases were run on a single NVIDIA RTX A6000 GPU. 


\subsection{Data-driven training task}

\subsubsection{Training on parametric PDE solutions}
\label{subsec:training_on_pde_solutions}
In this example, we aim to learn the solution to a time-dependent parametric PDE that models heat transfer inside the domain. The input parameters are spatial location $(x, y)$, time $t$, heat conductivity $k$, and source power $P$. The output is the temperature $u$. As a result, the function mapping of the parametric PDE can be written as $(x, y, k, P, t) \rightarrow u$.

The governing parametric PDE can be written as:
\begin{equation}
    \begin{cases}\dot{u}(\boldsymbol{x},t)+k\Delta u(\boldsymbol{x},t)=b(\boldsymbol{x}, P)&\mathrm{in}\quad\Omega_{\bm{x}}\otimes\Omega_t,\\u(\boldsymbol{x},t)|_{\partial\Omega}=0,\\u(\boldsymbol{x},0)=0, & \end{cases}
\label{spt_pde}
\end{equation}
defined in a spatial domain $\Omega_{\bm{x}}=[0,1]^2$, a temporal domain $\Omega_t=(0,0.04]$, and parametric domains $\Omega_k=[1,4]$ and $\Omega_P=[100,200]$.

\begin{figure}[h]
\centering
\includegraphics[width=0.6\linewidth]{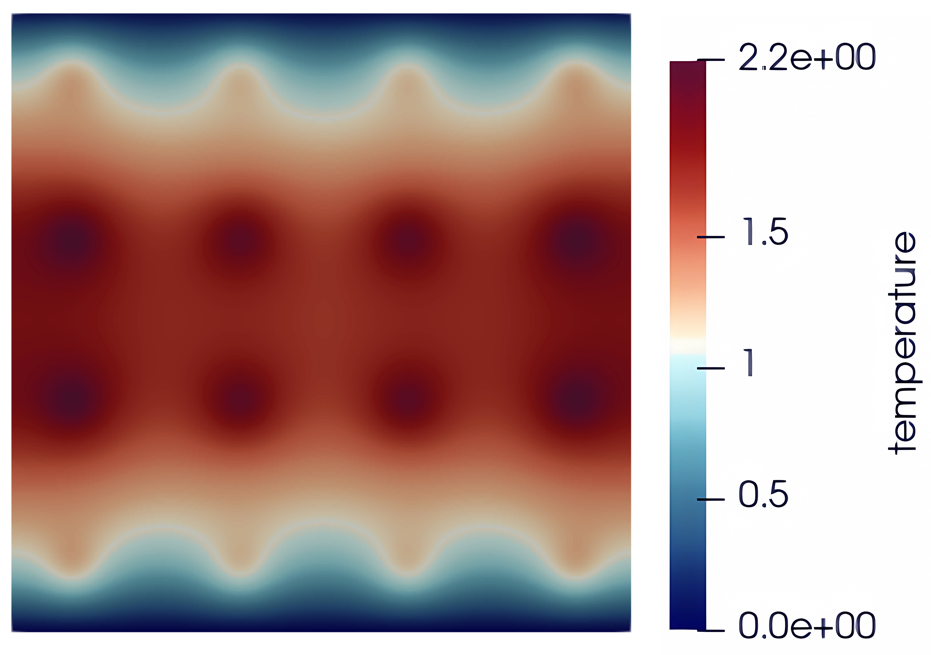}
\caption{Problem statement}
\label{problem_statement}
\end{figure}

We use the Gaussian source function to model the right-hand side (RHS) source term $b(\boldsymbol{x},P)$:
\begin{equation}
    b(\boldsymbol{x}, P)=\Sigma_{i=1}^{n_s}P\exp\left(-\frac{2((x - x_i)^2+(y - y_i)^2)}{r_0^2}\right)
\label{pde}
\end{equation}

where $r_0=0.05$ is the standard deviation that characterizes the width of the source function and $(x_i, y_i)$ is the $i$-th source center. We used $n_s=16$ in this example as shown in Fig. \ref{problem_statement}.

The simulation data are generated by running finite element analysis (FEA) with different parameters: $k$ and $P$. We compare 4 different methods: MLP, SIREN, KAN, INN-TD. The results are summarized in Table. \ref{tab:model_comparison_PDE_trainer} where the root mean squared error (RMSE) is reported. Details on the model setup are provided in Appendix \ref{sec:append_model_setup}. We can see that INN-TD outperforms in the training and test sets. Note that for KAN, the 30\% and 100\% training data cases use a batch size of 256 to prevent out-of-memory issues, while the 10\% training data case employs full-batch optimization, resulting in a significantly lower error than the other two cases. Nevertheless, for all cases listed here, INN has higher accuracy compared to the other 3 different models.

\begin{table}[htbp]
\centering
\scriptsize 
\caption{Root Mean Square Error (RMSE) Comparison for Training and Test Sets}
\begin{tabular}{llcccc}
\toprule
\multirow{2}{*}{Dataset} & \multirow{2}{*}{Model} & \multicolumn{2}{c}{Training ($\times 10^{-3}$)} & \multicolumn{2}{c}{Test ($\times 10^{-3}$)} \\
\cmidrule(lr){3-4} \cmidrule(lr){5-6}
& & Mean & Std  & Mean  & Std  \\
\midrule
\multirow{4}{*}{Training 10\%} & MLP & 4.180 & 0.873 & 3.660 & 0.498 \\
& SIREN & 2.330 & 0.188 & 4.550 & 0.297 \\
& KAN & 4.600 & 0.243 & 4.840 & 0.323 \\
& INN-TD & \textbf{0.213} & 0.009 & \textbf{0.748} & 0.911 \\
\midrule
\multirow{4}{*}{Training 30\%} & MLP & 0.699 & 0.143 & 0.912 & 0.155 \\
& SIREN & 1.240 & 0.161 & 2.020 & 0.197 \\
& KAN & 4.710 & 1.120 & 4.490 & 0.996 \\
& INN-TD & \textbf{0.154} & 0.003 & \textbf{0.154} & 0.006 \\
\midrule
\multirow{4}{*}{Training 100\%} & MLP & 0.385 & 0.037 & 0.534 & 0.041 \\
& SIREN & 1.170 & 0.121 & 1.660 & 0.200 \\
& KAN & 4.590 & 0.231 & 4.260 & 0.087 \\
& INN-TD & \textbf{0.126} & 0.002 & \textbf{0.129} & 0.005 \\
\bottomrule
\end{tabular}
\label{tab:model_comparison_PDE_trainer}
\end{table}

\subsection{Data-free Solving task}
\label{sec:solving}
\subsubsection{Solving high-dimensional PDE}
\label{sec:high_pde}
In this example, we use INN-TD as a data-free solver to directly solve high-dimensional Poisson's equation with different source terms and compare its performance in terms of accuracy and speed against CP-PINN and KAN. Detailed model setups can be found in Appendix \ref{sec:append_model_setup}.

The multi-dimensional Poisson's equation is written as:

\begin{equation}
    \Delta u(x_1,x_2,...,x_D) = f(x_1,x_2,...,x_D)
    \label{poisson}
\end{equation}
where $\Delta (\cdot)$ is the Laplace operator which is defined as: $\Delta (\cdot):= \frac{\partial^2(\cdot)}{\partial x_1^2} + \frac{\partial^2(\cdot)}{\partial x_2^2}+,..., \frac{\partial^2(\cdot)}{\partial x_D^2}$; $D$ denotes the total number of dimensions. 

Point-wise relative L2 norm error is used to measure the accuracy since INN-TD is also compared with collocatio-based machine learning method:
\begin{equation}
    \epsilon=\left[\sum_{k=1}^{K}[\mathcal{J}u(\bm{x}_{k})-u^{exact}(\bm{x}_{k})]^{2}\right]^{\frac{1}{2}}/\left[\sum_{k=1}^{K}u^{exact}(\bm{x}_{k})^{2}\right]^{\frac{1}{2}}
\label{error}
\end{equation}
where $K$ is the total number of test data; $u^{exact}(\bm{x}_{k})$ is the analytical solution. We let $f(\bm{x}) = -\frac{\pi^2}{4}\sum_{i=1}^D\sin\left(\frac{\pi}{2}x_i\right)$ and the corresponding analytical solution is: 

\begin{equation}
    u^{exact}(\bm{x})=\sum_{i=1}^n\left(\sin\left(\frac{\pi}{2}x_i\right)\right)
\label{exact}
\end{equation}

\begin{figure}
    \centering
    \begin{subfigure}[b]{0.48\linewidth}
        \centering
        \includegraphics[width=\linewidth]{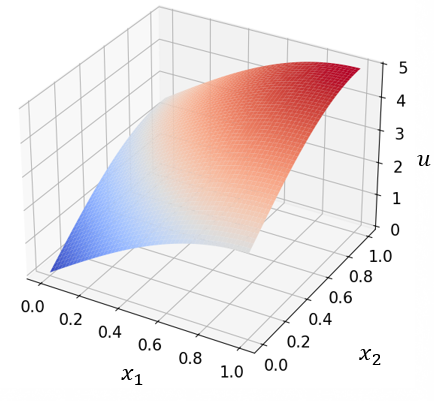}
        \caption{Domain size $\bm{x} \in [0, 1]^2$}
    \end{subfigure}
    \hfill
    \begin{subfigure}[b]{0.48\linewidth}
        \centering
        \includegraphics[width=\linewidth]{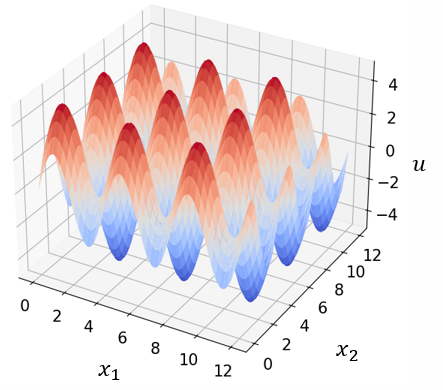}
        \caption{Domain size $\bm{x} \in [0, 12]^2$}
    \end{subfigure}
    \caption{Analytical solution for 2D cases.}
    \label{addfig}
\end{figure}

Two cases are examined here. In Case 1, the domain size is $[0, 1]^D$ , while Case 2 explores a larger domain size $[0, 12]^D$. The analytical solution is shown in Fig. \ref{addfig} for both cases when $D=2$. As evident from the figure, Case 2 is more challenging to solve due to the presence of higher-frequency signals in its analytical solution. It's known that vanilla MLP has spectral bias and thus poses challenges to learn Case 2 accurately and efficiently \cite{rahaman2019spectral}.

Table. \ref{add1} lists the detailed performance of each different method in Case 1. INN-TD stands out due to its better performance in terms of accuracy and computational efficiency. For instance, in the 2D setting, INN-TD achieves a low relative L2 norm error of $1.754 \times 10 ^{-8}$ in just 0.81 seconds, significantly outperforming other methods in both accuracy and speed. As the dimensionality increases to 5 and 10, INN-TD maintains its high accuracy with errors of $1.659 \times 10 ^{-8}$ and $1.238 \times 10 ^{-8}$, respectively, while still being computationally efficient. CP-PINN and KAN are able to tackle high-dimensional problems. However, since the total number of collocation points will increase exponentially with the increase of dimension, they still suffer from expensive  computational costs for high-dimensional problems. This demonstrates the robustness and scalability of the INN-TD method, making it a promising approach for solving high-dimensional PDEs.

\begin{table}[!hbt]
\caption{Comparison of different data-free solvers with domain size $[0, 1]^D$: INN-TD outperforms other models in both accuracy and speed. The mean values for error and wall time are reported.}
\scriptsize
\label{add1}
\begin{tabularx}{0.5\textwidth}{lXXXp{2cm}XX}
\hline
\multicolumn{1}{l}{\textbf{Model}} &
  \textbf{\# dim} &
  \textbf{\# collocation points in each dim} &
  \textbf{\# model parameters} &
  \textbf{Relative L2 norm error} &
  \textbf{GPU wall time (sec)} \\ \hline
\multirow{3}{*}{\textbf{INN-TD}}  & 2  & -   & 256   & $1.754 \times 10^{-8}$ & $0.81$  \\
                                  & 5  & -   & $1,\!600$ & $1.659 \times 10^{-8}$ & $6.88$ \\
                                  & 10 & -   & $6,\!400$ & $1.238 \times 10^{-8}$ & $57.43$  \\ \hline
\multirow{3}{*}{\textbf{CP-PINN}} & 2  & 32  & 96    & $4.63 \times 10^{-3}$ & $60.68$   \\
                                  & 2  & 128 & 96    & $2.64 \times 10^{-3}$ & $169.2$   \\
                                  & 5  & 32  & $1,\!200$ & $5.90 \times 10^{-3}$ & $2,130$  \\ \hline
\multirow{3}{*}{\textbf{KAN}}     & 2  & 32  & 348   & $9.44 \times 10^{-5}$ & $56.16$   \\
                                  & 2  & 128 & 348   & $4.84 \times 10^{-5}$ & $58.88$    \\
                                  & 5  & 12  & 624   & $1.68 \times 10^{-4}$ & $842.1$  \\ \hline
\end{tabularx}%
\end{table}

Table. \ref{add12} provides a performance comparison of different models in Case 2. It is evident that the accuracy of both CP-PINN and KAN significantly decrease compared to Case 1, due to the increase in domain size. Despite utilizing more collocation points, the relative L2 norm error for these two methods does not converge to a smaller value, indicating possible limitations in applying them to industrial-level problems requiring high accuracy and solver convergence. In contrast, INN-TD demonstrates significantly better accuracy, outperforming other methods by orders of magnitude while requiring much less computational time.

\begin{table}[!hbt]
\caption{Comparison of different data-free solvers with domain size $[0, 12]^D$: INN-TD outperforms other models in both accuracy and speed. The mean values for error and wall time are reported.}
\scriptsize
\label{add12}
\begin{tabularx}{0.5\textwidth}{lXXXp{2cm}XX}
\hline
\textbf{Model} & \textbf{Dim} & \textbf{\# collocation points in each dim} & \textbf{\# model parameters} & \textbf{Relative L2 norm error} & \textbf{GPU wall time (sec)} \\ \hline
\multirow{4}{*}{\textbf{INN-TD}}  & 2  & -     & 256     & $4.77 \times 10^{-4}$ & $0.22$ \\
                                  & 5  & -     & $1,\!600$   & $3.97 \times 10^{-4}$ & $7.04$  \\
                                  & 10 & -     & $6,\!400$   & $3.71 \times 10^{-4}$ & $59.75$ \\ \hline
\multirow{3}{*}{\textbf{CP-PINN}} & 2  & 32    & 216     & $2.5 \times 10^{-2}$ & $87.6$  \\
                                  & 2  & $1,\!024$ & 216     & $2.1 \times 10^{-2}$ & $261.6$ \\
                                  & 2  & $2,\!048$ & 296     & $3.2 \times 10^{-2}$ & $292$ \\ \hline
\multirow{4}{*}{\textbf{KAN}}     & 2  & 32    & 348     & $7.8 \times 10^{-1}$ & $60.68$  \\
                                  & 2  & 128   & 348     & $8.0 \times 10^{-1}$ & $72.6$  \\
                                  & 2  & 128   & $3,\!188$   & $2.8 \times 10^{-1}$ & $119.7$ \\
\hline
\end{tabularx}%
\end{table}

\begin{figure}[!hbt]
\centering
\includegraphics[width=0.75\linewidth]{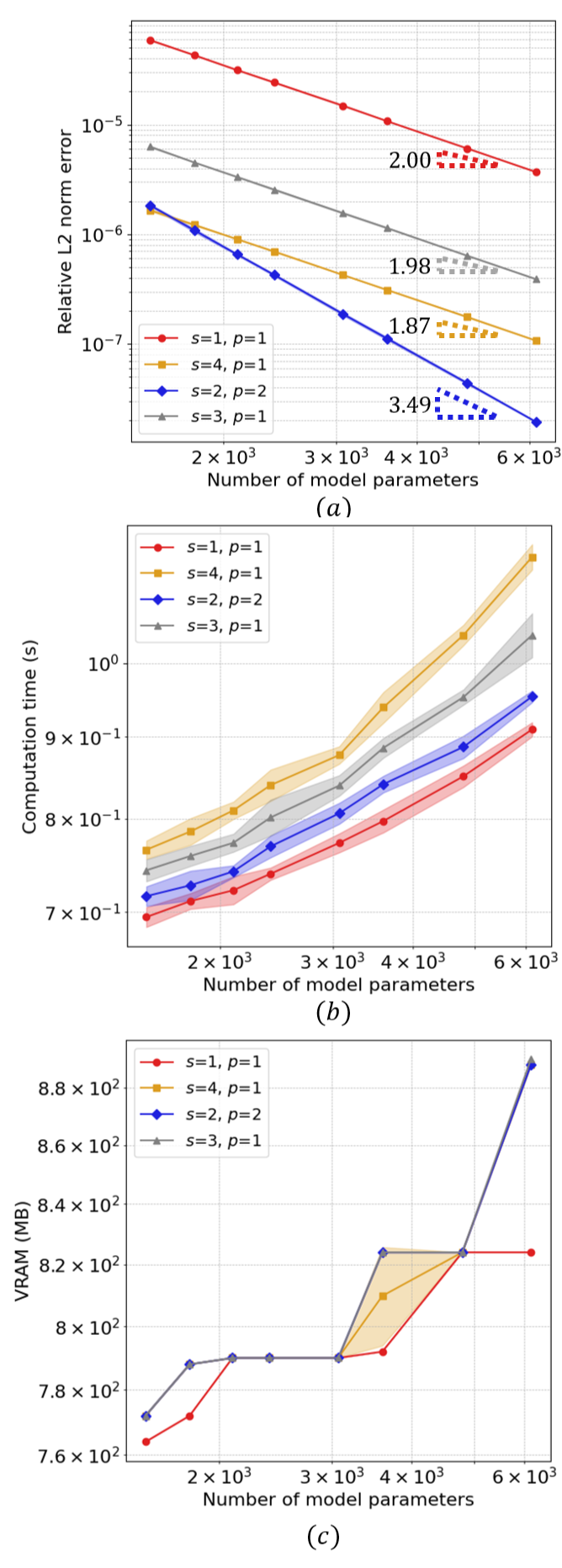}
\caption{Solving the 2D Poisson's Eq. using INN-TD with different number of model parameters and hyperparameters $s$ and $p$. $a$=20 is used for all cases. The statistics are obtained by repeating each case for 10 times and the shaded area represents 1 standard deviation (a) convergence of INN-TD using different $s$ and $p$ (b) statistics of total computational time (c) statistics of GPU VRAM usage}
\label{converge}
\end{figure}

Furthermore, we evaluate the convergence of the INN-TD data-free solver for Case 1. As illustrated in Fig.\ref{converge} (a), INN-TD demonstrates favorable convergence properties, with reduced error achieved as the model complexity increases. This convergence property enhances confidence in the model's accuracy, enabling the design of the network to align with the anticipated precision.

In addition to this example, three more examples are included in Appendix \ref{sec:pde_more} to further demonstrate the efficiency and accuracy of INN-TD as a data-free solver for tackling large-scale high-dimensional PDEs. In particular, we show INN-TD doesn't exhibit the failure modes of PINN when solving the Helmholtz equation \cite{wang2021understanding}.

\subsubsection{Solving general space-parameter-time (S-P-T) PDE}
\label{spt}
A distinct advantage of machine learning-based data-free solvers is their ability to naturally handle parametric PDEs by treating parametric variables as inputs. In contrast, standard numerical solvers like FEM must rerun simulations for each different set of parametric inputs. In this example, we demonstrate that INN-TD effectively handles parametric PDEs by utilizing a space-parameter-time (S-P-T) interpolation. We consider solving the time-dependent parametric PDE described in Eq. \ref{spt_pde}, treating it as a 3D spatial problem $\bm{x}_s = (x, y, z)$ with 2D parametric inputs $\bm{x}_p = (k, P)$ and 1D temporal input $x_t = t$.

\begin{figure}[!hbt]
\centering
\includegraphics[width=0.65\linewidth]{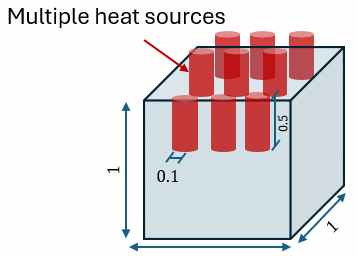}
\caption{S-P-T data-free solver problem}
\label{problem_statement3d}
\end{figure}

The new source function can be written as:
\begin{equation}
    b(\boldsymbol{x}, P)=\Sigma_{i=1}^{n_s}P\exp\left(-\frac{2((x - x_i)^2+(y - y_i)^2)}{r_0^2}\right) \mathbbm{1}_{z \geq d_0}
\label{depth_source}
\end{equation}

where $d_0$ is the depth of the source and $d_0 = 0.5$ in the current example; $\mathbbm{1}_{z \geq d_0}$ is the indicator function where $\mathbbm{1}_{z \geq d_0}$ equals 1 if $z\geq d_0$ or 0 if $z < d_0$. Therefore, this PDE can be interpreted as modeling heat transfer with multiple fixed heat sources, each characterized by a radius $r_0$ and a penetration depth $d_0$, as shown in Fig. \ref{problem_statement3d}.

\begin{figure}[!hbt]
\centering
\includegraphics[width=\linewidth]{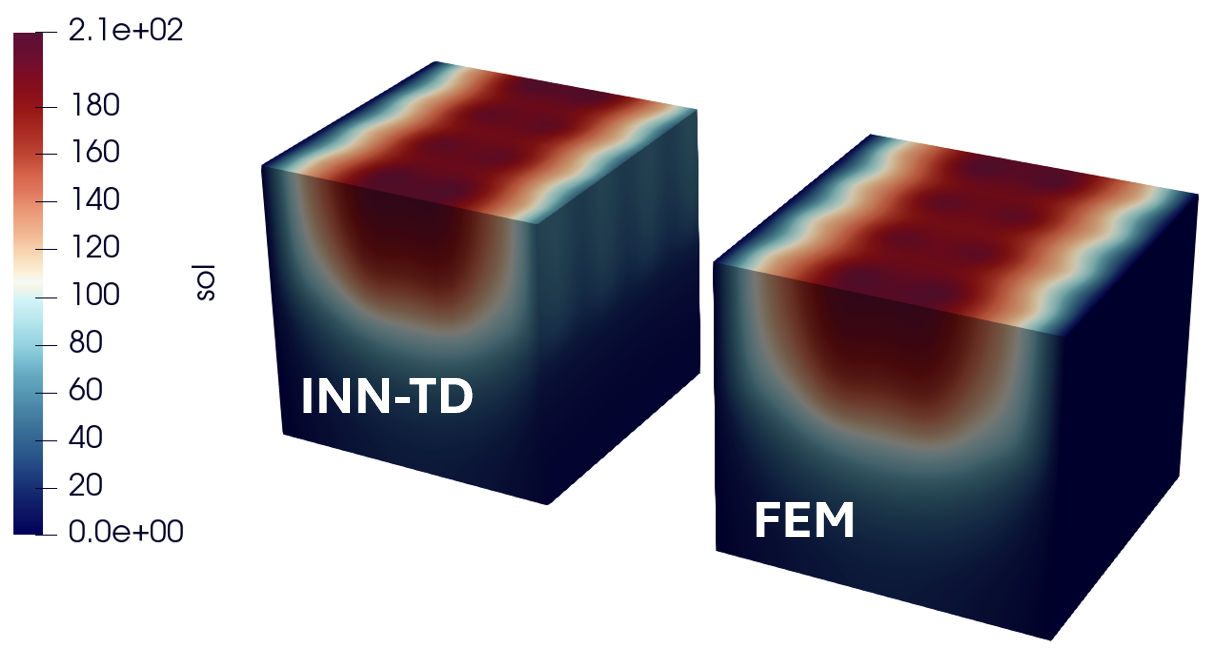}
\caption{Comparison of INN-TD and FEM solution for $k=1.03, P=101, t=0.95$}
\label{spt_compare}
\end{figure}

This problem is solved using INN-TD using the algorithm listed in Algorithm \ref{algo:solver} by discretizing each input dimension with 101 grid points and utilizing 100 modes. The total solving time takes 155 s on a single NVIDIA RTX A6000 GPU. Fig. \ref{spt_compare} compares the INN-TD solution with FEM when $k=1.03, P=101, t=0.95$.

To quantitatively assess the accuracy of the solution obtained from INN-TD, we measured the accuracy of INN-TD against the high-fidelity simulation by running implicit FEM with different realizations of parametric inputs $(k, P)$. The relative L2 norm error for INN-TD is $3.38 \times 10^{-3}$. Details about the error computation can be found in Appendix \ref{sec:spt}. 

More data-free solving problems can be found in Appendix \ref{sec:pde_more}.

\subsection{Inverse optimization task}
In the previous section, we have shown INN-TD can effectively capture complex S-P-T function mapping for both data-driven training and data-free solving tasks. Consequently, for the inverse problem, we first train (solve) the function mapping from S-P-T continuum to output using the data-driven trainer (data-free solver). Then we recover the unknown parametric coefficients by minimizing the difference between prediction and measurements using automatic differentiation. 

As an example, we use the same PDE as in section \ref{spt} to illustrate the performance of INN-TD for the inverse problem. In this task, we aim to find the optimal parametric inputs $(k, P)$ to obtain the target spatial-temporal field $u^{*}(\bm{x_s}, x_t)$.

We adopt the following 2 metrics to analyze the accuracy. The error of identified parametric input is defined as:
\begin{equation}
    \epsilon_{x_p} = \frac{|x_p - x_p^{*}|}{|x_p^{*}|}
\end{equation}
The error of the prediction is defined as the relative L2 norm error between estimated output and target output.

\begin{equation}
    \epsilon_{L2}=\frac{\|\mathcal{J}u(\bm{x_s}, x_t; \bm{x_p}) - u^{*}(\bm{x_s}, x_t)\|_2}{\|u^{*}(\bm{x_s}, x_t)\|_2}
\end{equation}

The error metrics of this example are summarized in Table. \ref{tab:opt_errors}. As shown in the table, INN-TD effectively recovers the optimal parametric inputs with high accuracy. Details on the model setups can be found in Appendix \ref{sec:inverse}.
\begin{table}[!hbt]
\centering
\small
\caption{Error for inverse optimization task.}
\begin{tabularx}{0.5\textwidth}{lXX}
\hline
\textbf{Error Type} & \textbf{Mean} & \textbf{Standard Deviation} \\
\hline
$\epsilon_{L_2}$ & $2.18\times 10^{-3}$ & $5.64 \times 10^{-5}$ \\

$\epsilon_{k}$  & $2.76\times 10^{-3}$ & $4.64\times 10^{-4}$ \\

$\epsilon_{P}$  & $2.62\times 10^{-3}$ & $3.11\times 10^{-4}$ \\
\hline
\end{tabularx}

\label{tab:opt_errors}
\end{table}


\section{Conclusion}
In this paper, we introduced INN-TD as an efficient and accurate function approximation for large-scale, high-dimensional problems governed by PDEs. By combining the strengths of machine learning techniques and classical numerical algorithms, INN-TD demonstrated high accuracy in data-driven training, data-free solving, and inverse optimization tasks for problems where the underlying physics are described by PDEs. Particularly in the solving task, INN-TD achieves orders of magnitude better accuracy compared to other machine learning models. Convergence properties have been observed where more model parameters result in higher accuracy. Furthermore, INN-TD efficiently scales to higher-dimensional problems with extremely fast solving speeds. Lastly, due to its interpretable nature, we can tailor the structure of INN-TD according to the required resolution in physical problems. In conclusion, by unifying training, solving, and inverse optimization within a single framework, INN-TD offers a seamless approach to addressing challenging, large-scale physical problems that demand high precision and rapid solution speeds.

\section*{Impact Statement}

INN-TD lies at the intersection of artificial intelligence (AI) and traditional simulation approaches. By embedding inductive biases inspired by numerical analysis directly into the network architecture, INN-TD effectively tackles the challenges of low accuracy and high computational cost associated with current deep learning models when applied to high-dimensional, large-scale physical problems governed by PDEs.


\bibliography{example_paper}
\bibliographystyle{icml2025}

\newpage
\appendix
\onecolumn
\section{C-HiDeNN interpolation theory}
\label{sec:append_chidenn}
\begin{figure}[!hbt]
\centering
\includegraphics[width=0.95\linewidth]{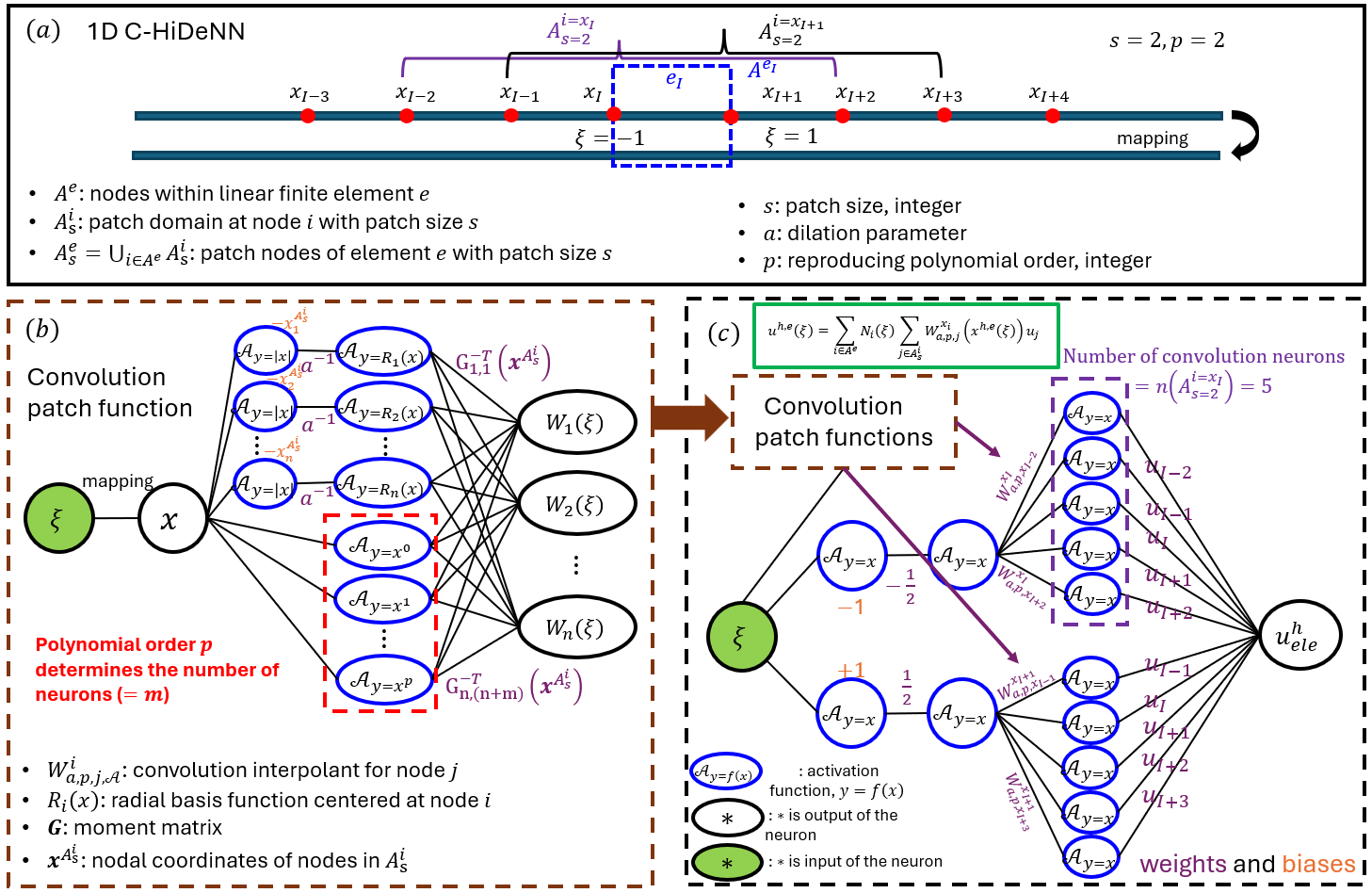}
\caption{(a) Convolution patch in 1D C-HiDeNN shape function (b) Construction of convolution patch function (c) C-HiDeNN shape function as MLP with 3 hidden layers. This plot is adopted from \cite{park2024engineering}.}
\label{chidenn_fig}
\end{figure}

Convolutional-Hierarchical Deep Neural Network
(C-HiDeNN)  interpolation theory synergizes the advantages of finite element interpolation, mesh-free interpolation, and machine learning optimization. \cite{lu2023convolution, park2023convolution}. A 1D C-HiDeNN interpolation can be written in the algebraic form as
\begin{equation}
    \mathcal{J}u(x)=\sum_{i\in A^e}N_i(x)\sum_{j\in A_s^i}\mathcal{W}_i^{(j)}(x; s,a,p)u_j=\sum_{k\in A_s^e}\widetilde{N}_k(x;s,a,p)u_k=\widetilde{\bm{N}}(x;s,a,p)\boldsymbol{u}
    \label{chidenn}
\end{equation}

 where $N_i(x)$ is the standard finite element basis function at node $i$; $W_i^{(j)}{(x; s,a,p)}$ is the convolution patch function at node $j$
defined on the support domain $A_s^i$. The double summation can be combined to a single summation over elemental nodal patches defined as $A_s^e = \bigcup_{i \in A^e} A_s^i$. We can also interpret Eq. \ref{chidenn} as a partially connected MLP, as shown in Fig. \ref{chidenn_fig}.

 In Fig. \ref{chidenn_fig}, the convolution patch function $W_i^{(j)}{(x; s,a,p)}$ is controlled by three hyperparameters: patch size $s$ that controls nodal connectivity, dilation parameter $a$ that normalizes distances between patch nodes, and reproducing order $p$ that defines types/orders of activation functions to be reproduced by the patch functions. C-HiDeNN can adapt to these hyperparameters node by node using optimization, rendering an adaptable functional space without altering the number of nodes or layers.

 Unlike the traditional black-box MLP, C-HiDeNN can be interpreted as a tailored MLP structure that preserves several desired properties (inductive biases) that are particularly important in numerical analysis: locality, Kronecker delta property \cite{hughes2003finite}, partition of unity \cite{melenk1996partition}, and easy for integration \cite{hughes2003finite}. These attributes enable C-HiDeNN to efficiently model complex physical problems with high accuracy.

\begin{table}[!htb]
\centering
\caption{Comparison of MLP and C-HiDeNN}
\begin{tabularx}{0.8\textwidth}{|X|X|X|}
\hline
\textbf{Properties}         & \textbf{MLP}                                              & \textbf{C-HiDeNN}                                                    \\ \hline
Boundary/initial condition & Penalty term in the loss function \cite{raissi2019physics}        & Automatic satisfaction due to Kronecker delta \cite{lu2023convolution}      \\ \hline
Convergence and stability  & Stochastic and not guaranteed \cite{colbrook2022difficulty} & Shown for different PDEs \cite{guo2024convolutional}                           \\ \hline
Numerical integration      & Quasi-Monte Carlo integration \cite{kharazmi2021hp}                    & Gaussian quadrature \cite{hughes2003finite}                                      \\ \hline
Interpretability           & Black-box model                                  & Fully interpretable                                                   \\ \hline 
\end{tabularx}
\end{table}

\section{Interpretability of INN-TD}
\label{sec:Inter}
\begin{figure}[h]
\centering
\includegraphics[width=0.9\linewidth]{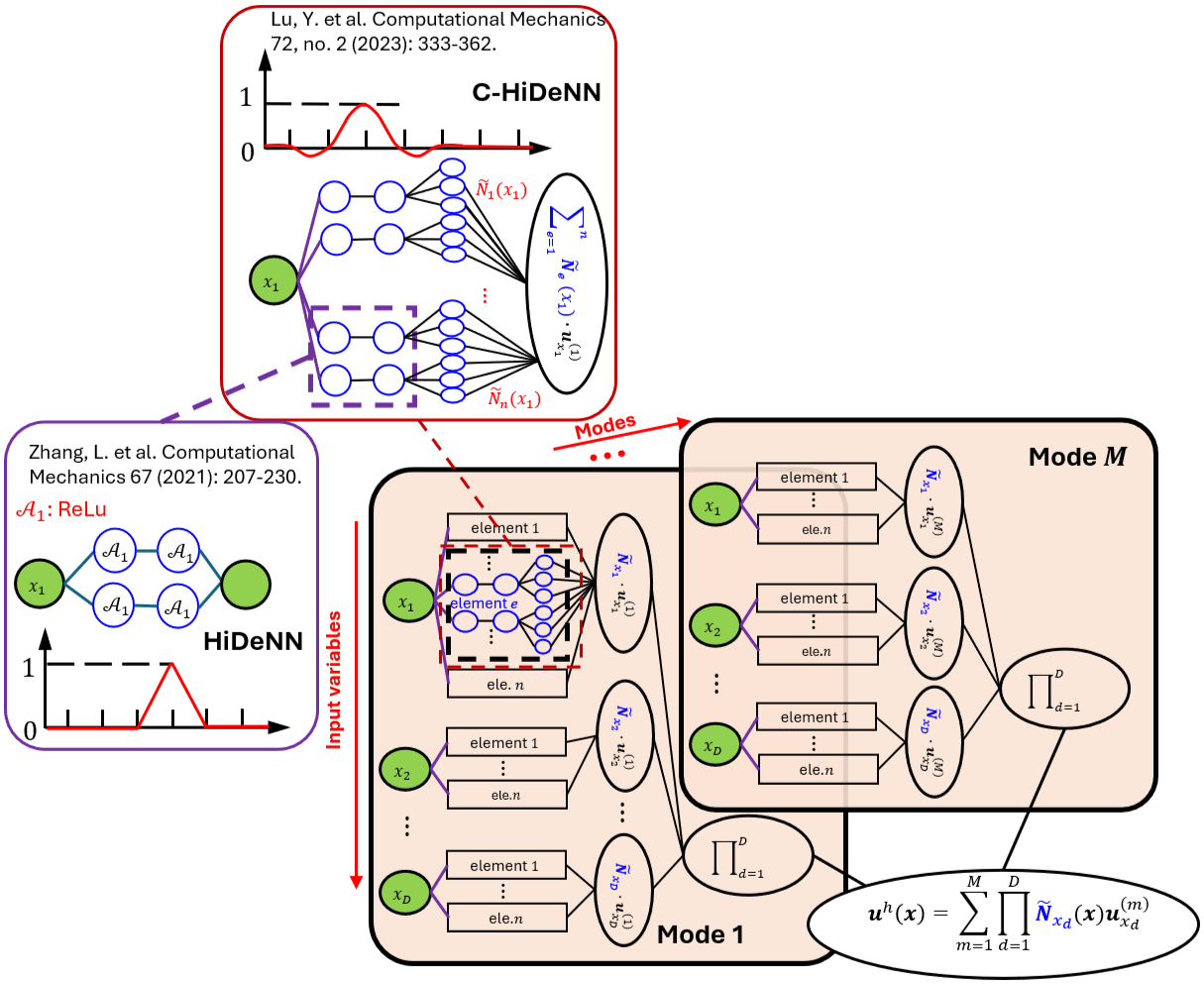}
\caption{Interpretable structure of INN-TD}
\label{inn-td}
\end{figure}

Unlike most black-box machine learning models, INN-TD has a fully interpretable architecture. As shown in Fig. \ref{inn-td}, INN-TD can be interpreted as a specially pruned MLP where the first 2 hidden layers represent locally supported linear finite element basis functions. The 3rd hidden layer reconstructs the C-HiDeNN interpolation function with higher-order smoothness which is controlled by hyperparameter patch size $s$ and reproducing polynomial order $p$. The 4th hidden layer leverages tensor decomposition to counter the curse of dimensionality. As a result, INN-TD's learning parameters are interpretable nodal values of TD components, in contrast to the opaque weights and biases of neural networks. 

To demonstrate the benefits of INN's interpretability, we solve the following 2D Poisson's equation in $\Omega =[0,100]^2$ with a local source function.
\begin{equation}\left(\frac{\partial^2}{\partial x_1^2}+\frac{\partial^2}{\partial x_2^2}\right)u=f(x,y)\end{equation}

The local Gaussian source function $f(x,y)$ is defined as:
\begin{equation}\begin{gathered}f(x,y) = 0.064(x-40)^2e^{-0.04(x-44)^2}e^{-0.04(y-75)^2}+0.064(x-20)^2e^{-0.04(x-20)^2}e^{-0.04(y-25)^2}\\+0.064(y-15)^2e^{-0.04(x-46)^2}e^{-0.04(y-55)^2}+0.064(y-25)^2e^{-0.04(x-26)^2}e^{-0.04(y-25)^2}\\-1.6e^{-0.04(x-20)^2}e^{-0.04(y-25)^2}-1.6e^{-0.04(x-40)^2}e^{-0.04(y-55)^2}\end{gathered}\end{equation}

The exact solution to this problem is: 
\begin{equation}u^{ex}(x,y)=10e^{-\frac{(x-20)^2}{25}}e^{-\frac{(y-25)^2}{25}}+10e^{-\frac{(x-60)^2}{25}}e^{-\frac{(y-75)^2}{25}}\end{equation}

The Dirichlet boundary condition is $u|_{\partial\Omega}=u^{ex}(x,y)|_{\partial\Omega}$.

\begin{figure}[h]
\centering
\includegraphics[width=1.0\linewidth]{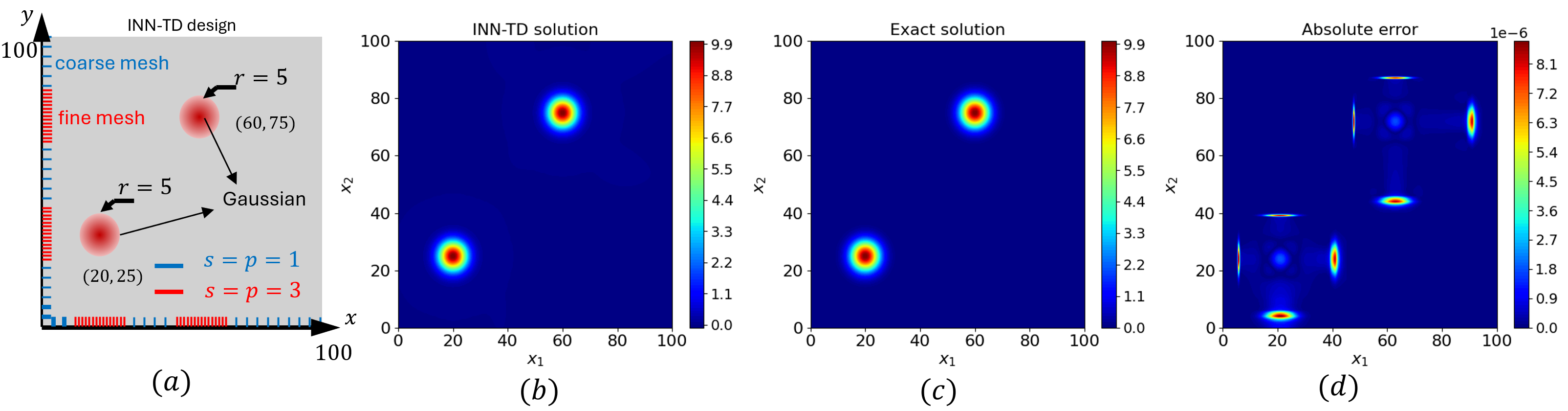}
\caption{Interpretable design of INN-TD: (a) 1D meshes used in INN-TD are only refined at the location where nonlinearity happens with larger $s$ and $p$ used to achieve higher-order smoothness (b) INN-TD solution (c) Exact solution (d) point-wise absolute error}
\label{interp_example}
\end{figure}

To efficiently solve this problem, we leverage the interpretable locally supported basis function in INN-TD. As can be seen from Fig. \ref{interp_example} (a), a nonuniform mesh is used to accommodate the local source function: a fine mesh is used for the source function region, whereas a coarse mesh is used for other regions. Moreover, since the nonlinearity of the solution is expected to be localized near the source function region, larger $s$ and $p$ can be used for these local regions to improve the accuracy. The absolute pointwise error with respect to the exact solution is plotted in Fig. \ref{interp_example} (d) where the order is around $10^{-6}$.

\section{INN-TD trainer}
\label{sec:append_trainer}
Two different training schemes can be used for INN-TD trainer. The first method is based on the boosting algorithm. The core idea behind boosting is to convert a set of weak learners into a strong learner. A weak learner is a model that performs slightly better than random guessing. The boosting process involves sequentially training multiple weak learners, each trying to correct the errors made by its predecessor.

We define the weak learner for mode $m$ as $\mathcal{J}^{(m)}\bm{y(x; U}^{(m)}\bm{)}$, where $\bm{U}^{(m)} = [\bm{u_1^{(m)}, u_2^{(m)}, ..., u_D^{(m)}}]$ is the model parameter for mode $m$. The strong learner $\mathcal{J}\bm{{y(x; U)}}$ can be obtained by adding weak learners for each mode, and we define $\bm{U = [U^{(1)}, U^{(2)}, ..., U^{(M)}]}$.
\begin{equation}
    \mathcal{J}\bm{{y(x; U)}} = \Sigma_{m=1}^{M} \mathcal{J}^{(m)}\bm{y(x; U}^{(m)}\bm{)}
    \label{allatonce}
\end{equation}

The loss of model prediction and training data $\bm{(x^*, y^*)}$ is defined using the mean squared error:
\begin{equation}
    \mathcal{L} = \frac{1}{K}\sum_k(\mathcal{J}\bm{{y(x_k^*; U)}}-\bm{y_k}^*)^2.
    \label{train_boost0}
\end{equation}
Assuming the previous $M-1$ weak learners have been obtained, and we aim to learn the new weak learner $\mathcal{J}^{(M)}\bm{y(x; U}^{(M)}\bm{)}$,  Eq. \ref{train_boost0} can be written as:
\begin{equation}
    \mathcal{L} = \frac{1}{K}\sum_k(\Sigma_{m=1}^{M-1} \mathcal{J}^{(m)}\bm{y(x_k^*; U}^{(m)}\bm{)} + \mathcal{J}^{(M)}\bm{y(x_k^*; U}^{(M)}\bm{)} -\bm{y_k}^*)^2.
    \label{train_boost1}
\end{equation}
Consequently, the goal is to learn to model parameters $\bm{U}^{(M)}$ for new weak learner $\mathcal{J}^{(M)}\bm{y(x; U}^{(M)}\bm{)}$ using optimization schemes.
\begin{equation}
    \operatorname*{argmin}_{\bm{U}^{(M)}} \mathcal{L} = \frac{1}{K}\sum_k(\Sigma_{m=1}^{M-1} \mathcal{J}^{(m)}\bm{y(x_k^*; U}^{(m)}\bm{)} + \mathcal{J}^{(M)}\bm{y(x_k^*; U}^{(M)}\bm{)} -\bm{y_k}^*)^2.
    \label{train_boost}
\end{equation}

The complete boosting algorithm for INN-TD trainer is shown in algorithm \ref{bst}.

\begin{algorithm}[!htb]
   \caption{INN-TD trainer: boosting}
   \label{bst}
\begin{algorithmic}
    \STATE Define maximum number of weak learners (modes) $M$, maximum epoch number epoch${_{max}}$, grid for each dimension $\bm{x}_d$
    \STATE Initialize model  $\mathcal{J}{\bm{y(x}})= 0$
   \FOR {(epoch loop) epoch $ = 1$ {\bfseries to} epoch$_{max}$}
   \FOR {(boosting loop) $m = 1$ {\bfseries to} $M_{max}$}
   \STATE Initialize solution vector $\bm{U}^{(m)}$ for the  ${m}$-th weak learner.
   \STATE Optimize Eq. \ref{train_boost} using optimization schemes such as Adam
   \STATE Update learner:  $\mathcal{J}{\bm{y(x}})= \mathcal{J}{\bm{y(x}}) + \mathcal{J}^{(m)}\bm{y(x; U}^{(m)}\bm{)}$
    \STATE Check loss
   \ENDFOR
   \ENDFOR
\end{algorithmic}
\end{algorithm}

Instead of gradually enhancing the accuracy of the model by adding weak learners, one can also predefine the total number of modes $M$ required in the model and optimize the unknown parameters in the model in an all-at-once fashion as shown in Eq. \ref{allatonce}. Consequently, we treat the total number of modes as one additional hyperparameter. The all-at-once training strategy has been shown to generally lead to better accuracy compared to the boosting strategy when the total number of modes $M$ is the same.

The complete all-at-once algorithm for INN-TD trainer is shown in algorithm \ref{ato}. We use this algorithm for all of the training experiments presented in the paper. 
\begin{algorithm}[!htb]
   \caption{INN-TD trainer: all-at-once}
   \label{ato}
\begin{algorithmic}
    \STATE Define total number of modes $M$, maximum epoch number epoch${_{max}}$, grid for each dimension $\bm{x}_d$
   \STATE Initialize model parameters $\bm{U = [U^{(1)}, U^{(2)}, ..., U^{(M)}]}$.
   \FOR {(epoch loop) epoch $ = 1$ {\bfseries to} epoch$_{max}$}
   \STATE Optimize Eq. \ref{allatonce} using optimization schemes such as Adam.
   \STATE Check loss
   \ENDFOR
\end{algorithmic}
\end{algorithm}

\section{INN-TD solver}
\label{sec:append_solver}
Boosting algorithm is adopted for INN-TD solver. Such method is known as proper generalized decomposition (PGD) in the field of computational mechanics \cite{li2023convolution}. A similar all-at-once approach named a priori tensor decomposition is also available \cite{guo2024convolutional}. 
Different from the training approach where the loss is computed in a point-wise manner, we compute the PDE loss in a weighted-sum residual format. Without loss of generality, we focus on the discussion of the scalar solution field. Vector and higher-order tensorial solution fields can be treated in a similar way.
Assume PDE can be written as:
\begin{equation}
  \mathcal{L}u(\bm{x}) = 0    
\end{equation}
where $\mathcal{L}$ is a general (nonlinear) differential operator, $u(\bm{x})$ is the $D$-dimensional PDE solution depending on independent variables $\bm{x} = (x_1, x_2, ..., x_D)$.  
The INN-TD approximation to the solution can be written as:
\begin{equation}
    \mathcal{J}u(\bm{x})=\sum_{m=1}^{M}\widetilde{\boldsymbol{N}}_{1}(x_1)\boldsymbol{u}_{x_1}^{(m)}\cdot\widetilde{\boldsymbol{N}}_2(x_2)\boldsymbol{u}_{x_2}^{(m)}\cdot... \cdot\widetilde{\boldsymbol{N}}_D(x_D)\boldsymbol{u}_{x_D}^{(m)}
\end{equation}
In the context of solutions to time-dependent parametric partial differential equations (PDEs), the independent variables $(x_1, x_2, \ldots, x_D)$ can be divided into three distinct categories: spatial variables $\bm{x_s}$, parametric variables $\bm{x_p}$, and temporal variables $x_t$. Spatial variables $\bm{x_s}$ define the spatial coordinates relevant to the problem. Parametric variables $\bm{x_p}$ serve as additional coordinates and can represent PDE coefficients, initial conditions, boundary conditions, and geometry descriptors. Temporal variable $x_t$ represents time.

The weighted-sum residual form is adopted to solve PDEs.

\begin{equation}
    \int\delta(\mathcal{J}u(\bm{x}))\mathcal{L}(\mathcal{J}u(\bm{x}))d\bm{x}=0
    \label{galerkin}
\end{equation}
where $\delta(\mathcal{J}u(\bm{x}))$ is the so-called weighting function. Depending on the form of the PDE, different choices of $\delta(\mathcal{J}u(\bm{x}))$ can be adopted. For instance, if the Dirac Delta function $\delta(\bm{x-x}_i)$ ($\delta$ here refers to the Dirac Delta function) is used for the weighting function where $\bm{x}_i$ is the collocation point, Eq. \ref{galerkin} becomes the differential (strong) form:   $\mathcal{L}u(\bm{x}) = 0$. In this paper, we adopt the Galerkin formulation, where the function space of the weighting function is the same as our approximation. Moreover, integration by parts can be used to alleviate the differentiability requirement on the approximated solution \cite{hughes2003finite}.

In the boosting solution scheme, the solution is obtained incrementally in a mode by mode fashion:

\begin{equation}
    \mathcal{J}u(\bm{x})=\sum_{m=1}^{M-1}\widetilde{\boldsymbol{N}}_{1}(x_1)\boldsymbol{u}_{x_1}^{(m)}\cdot\widetilde{\boldsymbol{N}}_2(x_2)\boldsymbol{u}_{x_2}^{(m)}\cdot... \cdot\widetilde{\boldsymbol{N}}_D(x_D)\boldsymbol{u}_{x_D}^{(m)}+\widetilde{\boldsymbol{N}}_1(x_1)\boldsymbol{u}_{x_1}\cdot\widetilde{\boldsymbol{N}}_2(x_2)\boldsymbol{u}_{x_2}\cdot...\cdot\widetilde{\boldsymbol{N}}_D(x_D)\boldsymbol{u}_{x_D}
    \label{trial}
\end{equation}
where we neglect the superscript $M$ for the last mode $M$ that we are solving.
As a result, the corresponding weighting function can be written as:
\begin{equation}
    \begin{aligned}\delta \mathcal Ju(\bm{x})&=\widetilde{\boldsymbol{N}}_{1}(x_1)\delta\boldsymbol{u}_{x_1}\cdot\widetilde{\boldsymbol{N}}_{2}(x_2)\boldsymbol{u}_{x_2}\cdot...\cdot\widetilde{\boldsymbol{N}}_{D}(x_D)\boldsymbol{u}_{x_D}+\\&\widetilde{\boldsymbol{N}}_{1}(x_1)\boldsymbol{u}_{x_1}\cdot\widetilde{\boldsymbol{N}}_{2}(x_2)\delta\boldsymbol{u}_{x_2}\cdot...\cdot\widetilde{\boldsymbol{N}}_{D}(x_D)\boldsymbol{u}_{x_D}+\\&\widetilde{\boldsymbol{N}}_{1}(x_1)\boldsymbol{u}_{x_1}\cdot\widetilde{\boldsymbol{N}}_{2}(x_2)\boldsymbol{u}_{x_2}\cdot...\cdot\widetilde{\boldsymbol{N}}_{D}(x_D)\boldsymbol{u}_{x_D}+\\&\widetilde{\boldsymbol{N}}_{1}(x_1)\boldsymbol{u}_{x_1}\cdot\widetilde{\boldsymbol{N}}_{2}(x_2)\boldsymbol{u}_{x_2}\cdot...\cdot\widetilde{\boldsymbol{N}}_{D}(x_D)\delta\boldsymbol{u}_{x_D}\end{aligned}
    \label{weighting}
\end{equation}

Based on Eq. \ref{weighting}, we further use subspace iteration to linearize the problem. In subspace iteration, we sequentially alternate 1 unknown variable while treating others as known constant values. In this case, the variation for other unknowns will be 0. Taking subspace iteration in $x_1$ as an example, Eq. \ref{weighting} becomes:

\begin{equation}
\delta_{x_1} \mathcal Ju(\bm{x})=\widetilde{\boldsymbol{N}}_{1}(x_1)\delta\boldsymbol{u}_{x_1}\cdot\widetilde{\boldsymbol{N}}_{2}(x_2)\boldsymbol{u}_{x_2}\cdot...\cdot\widetilde{\boldsymbol{N}}_{D}(x_D)\boldsymbol{u}_{x_D}
    \label{weighting_x}
\end{equation}

Plugging Eq. \ref{trial}, \ref{weighting_x} into Eq. \ref{galerkin} and integrating along each 1D dimension, the matrix equations for $x_1$ dimension can be derived:
\begin{equation}
    \bm{A}_{x_1} \bm{u}_{x_1} = \bm{Q}_{x_1}
    \label{td_linearx}
\end{equation}

where $\bm{u}_{x_1}$ is the solution vector for the current dimension at the current mode; $\bm{A}_{x_1}$ is a banded sparse matrix as in the standard finite element method thanks to locally supported basis functions; $\bm{Q}_{x_1}$ is a 1D vector.

Similarly, we can sequentially alternate other dimensions and obtain the corresponding linear system of equations:

\begin{equation}
    \bm{A}_{x_d} \bm{u}_{x_d} = \bm{Q}_{x_d}
    \label{td_linear}
\end{equation}

where $d = 1,..., D$; $\bm{u}_{x_d}$ is the solution vector for the current dimension at the current mode; $\bm{A}_{x_d}$ is a banded sparse matrix; $\bm{Q}_{x_d}$ is a 1D vector. As a result, Eq. \ref{td_linear} can be efficiently solved using sparse matrix solvers. We can keep iterating Eq. \ref{td_linear} until the variation of solution vector for each dimension is within tolerance. In practice, 3-5 iterations will yield good accuracy.

After the current mode is solved, we can introduce additional modes to further enhance the accuracy of the INN-TD model. This iterative process will be continued until the results meet the predefined accuracy criteria.

The complete boosting algorithm for INN-TD solver is shown in algorithm \ref{algo:solver}.
\begin{algorithm}[!htb]
   \caption{INN-TD solver: boosting}
   \label{algo:solver}
\begin{algorithmic}
    \STATE Define total number of modes $M$; maximum number of iteration $iter_{max}$; tolerance; grid for each dimension $\bm{x}_d$
    \STATE Initialize solution $\mathcal{J}u({\bm{x}})= 0$
   \STATE Initialize solution vector $\bm{u}^{(m)}_{x_d}$.
   \FOR {(boosting loop) $m = 1$ {\bfseries to} $M$} 
   \FOR {(subspace iteration loop) $iter=1$ {\bfseries to} $iter_{max}$}
   \FOR{(dimension loop) $d=1$ {\bfseries to} $D$}
   \STATE Update $\bm{A}_{x_d}$ and $\bm{Q}_{x_d}$
   \STATE Solve $\bm{A}_{x_d} \bm{u}^{(m)}_{x_d} = \bm{Q}_{x_d}$ using sparse matrix solver
   \ENDFOR
    \STATE Check convergence
   \ENDFOR
   \STATE $\mathcal{J}u({\bm{x}})= \mathcal{J} u({\bm{x}}) + \widetilde{\boldsymbol{N}}_{1}(x_1)\boldsymbol{u}_{x_1}^{(m)}\cdot\widetilde{\boldsymbol{N}}_2(x_2)\boldsymbol{u}_{x_2}^{(m)}\cdot... \cdot\widetilde{\boldsymbol{N}}_D(x_D)\boldsymbol{u}_{x_D}^{(m)} $
   \STATE Check Error
    \IF {Error $\leq$ Tolerance}
    \STATE Break
    \ENDIF
   \ENDFOR
\end{algorithmic}
\end{algorithm}


\section{Detailed model setups}
\label{sec:append_model_setup}

\subsection{Data-driven training examples}
\subsubsection{Model setups of training on parametric PDE solutions}
This section presents the hyperparameters for training models on the PDE dataset. For the MLP model, we set the initial learning rate to 0.001, the batch size to 128, and use the Adam optimizer. Training runs for a maximum of 500 epochs with early stopping based on the validation error. If the validation error increases for 10 consecutive epochs, the model training will be stopped. We also apply an adaptive learning rate scheduler, reducing the learning rate by 10\% after each epoch. The MLP model has 5 hidden layers, each having 50 neurons. 

For the SIREN model, we use almost the same hyperparameters as the MLP model. But we reduce the initial learning rate to 0.0001 to avoid loss explosion. We also increase the model complexity of the SIREN model. The model consists of 4 hidden layers, each containing 100 neurons.

For the KAN model, we set the layer widths to [5, 5, 5, 1] for the 10\% training data case, given that the input and output dimensions are 5 ($x, y, t, k, p$) and 1 ($u$), respectively. We use 10 grid intervals,third-order of piecewise polynomials, and the LBFGS optimizer with 2, 000 steps and full-batch training. For the 30\% and 100\% training data cases, full-batch training is infeasible due to the GPU memory limitation, so we set the batch size to 256. With a smaller batch size, we increase the layer widths to [5, 10, 10, 1] and extend the number of training steps to 3,000. Notably, training loss can easily explode to NaN and increasing batch size does not change the final error too much.

The INN-TD model uses 60 modes with 40 elements while the hyperparameters are: patch size $s=4$, dilation parameter $a=20$, and polynomial order $p=1$. Thus, there are a total of $5\times 60 \times 41=12,300$ trainable parameters. We use the batch size of 128 for 100 epochs with the same early stopping criteria adopted in MLP. The Adam optimizer is adopted with a learning rate of 0.0001.

\subsection{Data-free solver examples}
\label{sec:pde_more}
\subsubsection{High-dimensional PDE}
This subsection has detailed information discussed in section \ref{sec:high_pde}. Four different models, namely, INN-TD, PINN, CP-PINN, and KAN, were used as data-free solvers. The following high-dimensional PDE is solved.
\begin{equation}
    \Delta u(x_1,x_2,...,x_D) = f(x_1,x_2,...,x_D)
\end{equation}

In case 1, we adopt the following analytical solution:

\begin{equation}
    u^{exact}(\bm{x})=\sum_{i=1}^n\left(\sin\left(\frac{\pi}{2}x_i\right)\right)
\label{exact}
\end{equation}

The corresponding right-hand side (RHS) is: 
\begin{equation}
    f(\bm{x}) = -\frac{\pi^{2}}{4}\sum_{i=1}^{D}\sin\left(\frac{\pi}{2}x_{i}\right)
\end{equation}

Detailed performance comparison of different models is listed in the tables below.

\begin{table}[htbp]
\centering
\caption{Case 1: domain size $x\in[0,1]^D$ (each case is repeated 10 times to obtain the statistics)}
{\scriptsize 
\begin{tabular}{l c c c c c c c c}
\toprule
\textbf{Model} & \textbf{\# Dim.} & \textbf{\# colloc.} & \textbf{\# grid pts} & \textbf{\# modes} & \textbf{\# model} & \textbf{Mean rel. L2 norm} & \textbf{Mean GPU wall} & \textbf{Mean GPU} \\
& \textbf{($D$)} & \textbf{points} & \textbf{in each dim} & & \textbf{params.} & \textbf{error w/ std.} & \textbf{time w/ std.} & \textbf{VRAM usage} \\
& & & & & & & \textbf{(s)} & \textbf{w/ std. (MB)} \\
\midrule
\multirow{3}{*}{\parbox{2cm}{INN-TD \\ (4 iterations)}} & 2 & - & 32 & 4 & 256 & $1.754 \times 10^{-8} \pm 1 \times 10^{-12}$ & $0.81 \pm 0.10$ & $760 \pm 0$ \\
& 5 & - & 32 & 10 & 1,600 & $1.659 \times 10^{-8} \pm 1.2 \times 10^{-13}$ & $6.88 \pm 0.03$ & $760 \pm 0$ \\
& 10 & - & 32 & 10 & 6,400 & $1.238 \times 10^{-8} \pm 1.6 \times 10^{-12}$ & $57.43 \pm 0.41$ & $760 \pm 0$ \\
\midrule

\multirow{3}{*}{\parbox{2cm}{CP-PINN \\ (80\,K epochs)}} & 2 & $32^2$ & - & 4 & 96 & $4.63 \times 10^{-3} \pm 7.14 \times 10^{-4}$ & $60.68 \pm 0.53$ & $880 \pm 0$ \\
& 2 & $128^2$ & - & 4 & 96 & $2.64 \times 10^{-3} \pm 3.54 \times 10^{-4}$ & $169.2 \pm 4.83$ & $882 \pm 0$ \\
& 5 & $32^5$ & - & 10 & 1,200 & $5.90 \times 10^{-3} \pm 6.91 \times 10^{-4}$ & $2,130 \pm 18.6$ & $3018 \pm 0$ \\
\midrule
\multirow{3}{*}{\parbox{2cm}{KAN \\ (50 steps)}} & 2 & $32^2$ & 5 & - & 348 & $9.44 \times 10^{-5} \pm 8.37 \times 10^{-6}$ & $56.16 \pm 0.58$ & $866 \pm 0$ \\
& 2 & $128^2$ & 5 & - & 348 & $4.84 \times 10^{-5} \pm 8.95 \times 10^{-6}$ & $58.88 \pm 0.58$ & $1,160 \pm 0$ \\
& 5 & $12^5$ & 5 & - & 624 & $1.68 \times 10^{-4} \pm 2.45 \times 10^{-5}$ & $842.1 \pm 3.94$ & $9,988 \pm 0$ \\
\bottomrule
\end{tabular}
} 
\label{tab:case1}
\end{table}

\begin{table}[htbp]
\centering
\caption{Case 1: domain size $x \in [0,12]^D$ (each case is repeated 10 times to obtain the statistics)}
{\scriptsize 
\begin{tabular}{l c c c c c c c c}
\toprule
\textbf{Model} & \textbf{\# Dim.} & \textbf{\# colloc.} & \textbf{\# grid pts} & \textbf{\# modes} & \textbf{\# model} & \textbf{Mean rel. L2 norm} & \textbf{Mean GPU wall} & \textbf{Mean GPU} \\
& \textbf{($D$)} & \textbf{points} & \textbf{in each dim} & & \textbf{params.} & \textbf{error w/ std.} & \textbf{time w/ std.} & \textbf{VRAM usage} \\
& & & & & & & \textbf{(s)} & \textbf{w/ std. (MB)} \\
\midrule
\multirow{3}{*}{\parbox{2cm}{INN-TD \\ (4 iterations)}} & 2 & - & 32 & 4 & 256 & $4.77 \times 10^{-4} \pm 3.27 \times 10^{-8}$ & $0.22 \pm 0.006$ & $760 \pm 0$ \\
& 5 & - & 32 & 10 & 1,600 & $3.97 \times 10^{-4} \pm 4.62 \times 10^{-8}$ & $7.04 \pm 0.04$ & $760 \pm 0$ \\
& 10 & - & 32 & 10 & 6,400 & $3.71 \times 10^{-4} \pm 2.69 \times 10^{-9}$ & $59.75 \pm 0.39$ & $760 \pm 0$ \\
\midrule
\multirow{3}{*}{\parbox{2cm}{CP-PINN \\ (80\,K epochs)}} & 2 & $32^2$ & - & 4 & 216 & $0.025 \pm 0.004$ & $87.6 \pm 6.53$ & $882 \pm 0$ \\
& 2 & $1\text{,}024^2$ & - & 4 & 216 & $0.021 \pm 0.012$ & $261.6 \pm 7.79$ & $1\text{,}014 \pm 0$ \\
& 2 & $2\text{,}048^2$ & - & 4 & 296 & $0.032 \pm 0.007$ & $292 \pm 10.3$ & $1\text{,}014 \pm 0$ \\
\midrule
\multirow{3}{*}{\parbox{2cm}{KAN \\ (50 steps)}} & 2 & $32^2$ & 5 & - & 348 & $0.78 \pm 0.40$ & $60.68 \pm 0.53$ & $866 \pm 0$ \\
& 2 & $1\text{,}024^2$ & 5 & - & 348 & $0.80 \pm 0.27$ & $72.6 \pm 1.19$ & $19\text{,}948 \pm 0$ \\
& 2 & $128^2$ & 10 & - & 3,188 & $0.28 \pm 0.08$ & $119.7 \pm 2.05$ & $2\text{,}486 \pm 0$ \\
\bottomrule
\end{tabular}
} 
\label{tab:case12}
\end{table}

We also investigated the performance of four different models for the following analytical solution as in Case 2.
\begin{equation}
    u(\bm{x})=\prod_{d=1}^{D}\left(\sin(\pi x_{d})\right)
\end{equation}

\begin{figure}
    \centering
    \begin{subfigure}[b]{0.4\linewidth}
        \centering
        \includegraphics[width=\linewidth]{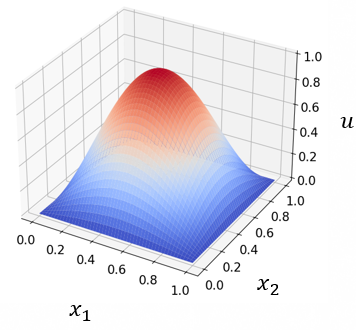}
        \caption{Domain size $\bm{x} \in [0, 1]^2$}
    \end{subfigure}
    \hfill
    \begin{subfigure}[b]{0.4\linewidth}
        \centering
        \includegraphics[width=\linewidth]{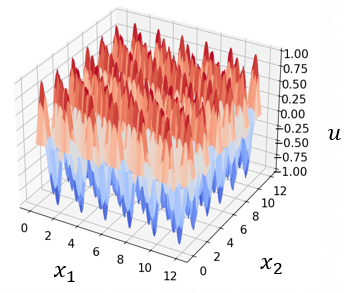}
        \caption{Domain size $\bm{x} \in [0, 12]^2$}
    \end{subfigure}
    \caption{Analytical solution for 2D cases.}
    \label{mulfig}
\end{figure}

For 2D cases, the analytical solutions are shown in Fig. \ref{mulfig}. 

The corresponding RHS can be written as:
\begin{equation}
    f(\bm{x}) = -D\pi^{2}\prod_{d=1}^{D}\left(\sin(\pi x_{d})\right)
\end{equation}
Similarly, we studied 2 subcases with different domain sizes. The detailed comparison is shown in the table below.

\begin{table}[htbp]
\centering
\caption{Case 2: domain size $x \in [0,1]^D$ (each case is repeated 10 times to obtain the statistics)}
{\scriptsize 
\begin{tabular}{l c c c c c c c c}
\toprule
\textbf{Model} & \textbf{\# Dim.} & \textbf{\# colloc.} & \textbf{\# grid pts} & \textbf{\# modes} & \textbf{\# model} & \textbf{Mean rel. L2 norm} & \textbf{Mean GPU wall} & \textbf{Mean GPU} \\
& \textbf{($D$)} & \textbf{points} & \textbf{in each dim} & & \textbf{params.} & \textbf{error w/ std.} & \textbf{time w/ std.} & \textbf{VRAM usage} \\
& & & & & & & \textbf{(s)} & \textbf{w/ std. (MB)} \\
\midrule
\multirow{3}{*}{\parbox{2cm}{INN-TD \\ (4 iterations)}} & 2 & - & 32 & 1 & 64 & $5.03 \times 10^{-7} \pm 1.31 \times 10^{-14}$ & $0.81 \pm 0.10$ & $760 \pm 0$ \\
& 5 & - & 32 & 1 & 160 & $1.23 \times 10^{-6} \pm 8.65 \times 10^{-14}$ & $1.01 \pm 0.02$ & $760 \pm 0$ \\
& 10 & - & 32 & 1 & 320 & $2.42 \times 10^{-6} \pm 8.1 \times 10^{-14}$ & $3.62 \pm 0.06$ & $764 \pm 0$ \\
\midrule
\multirow{3}{*}{\parbox{2cm}{CP-PINN \\ (80\,K epochs)}} & 2 & $32^2$ & - & 4 & 96 & $4.13 \times 10^{-3} \pm 1.00 \times 10^{-3}$ & $171.68 \pm 15.3$ & $880 \pm 0$ \\
& 2 & $128^2$ & - & 4 & 96 & $5.17 \times 10^{-3} \pm 1.38 \times 10^{-4}$ & $169.2 \pm 14.83$ & $882 \pm 0$ \\
& 5 & $32^5$ & - & 10 & 1,200 & $8.40 \times 10^{-3} \pm 1.45 \times 10^{-3}$ & $2\text{,}102 \pm 73.8$ & $3\text{,}018 \pm 0$ \\
\midrule
\multirow{3}{*}{\parbox{2cm}{KAN \\ (50 steps)}} & 2 & $32^2$ & 5 & - & 348 & $8.03 \times 10^{-4} \pm 6.90 \times 10^{-5}$ & $57.5 \pm 0.48$ & $866 \pm 0$ \\
& 2 & $128^2$ & 5 & - & 348 & $9.13 \times 10^{-4} \pm 1.90 \times 10^{-4}$ & $61.0 \pm 0.35$ & $1\text{,}160 \pm 0$ \\
& 5 & $12^5$ & 5 & - & 624 & $0.59 \pm 0.0054$ & $872.8 \pm 3.71$ & $9\text{,}988 \pm 0$ \\
\bottomrule
\end{tabular}
} 
\label{tab:case21}
\end{table}

\begin{table}[htbp]
\centering
\caption{Case 2: domain size $x \in [0,12]^D$ (each case is repeated 10 times to obtain the statistics)}
{\scriptsize 
\begin{tabular}{l c c c c c c c c}
\toprule
\textbf{Model} & \textbf{\# Dim.} & \textbf{\# colloc.} & \textbf{\# grid pts} & \textbf{\# modes} & \textbf{\# model} & \textbf{Mean rel. L2 norm} & \textbf{Mean GPU wall} & \textbf{Mean GPU} \\
& \textbf{($D$)} & \textbf{points} & \textbf{in each dim} & & \textbf{params.} & \textbf{error w/ std.} & \textbf{time w/ std.} & \textbf{VRAM usage} \\
& & & & & & & \textbf{(s)} & \textbf{w/ std. (MB)} \\
\midrule
\multirow{4}{*}{\parbox{2.3cm}{INN-TD \\ (4 iterations)}} & 2 & - & 32 & 1 & 64 & $1.56 \times 10^{-2} \pm 6.17 \times 10^{-14}$ & $0.77 \pm 0.30$ & $760 \pm 0$ \\
& 2 & - & 256 & 1 & 512 & $2.45 \times 10^{-6} \pm 1.37 \times 10^{-14}$ & $0.82 \pm 0.04$ & $760 \pm 0$ \\
& 5 & - & 32 & 1 & 160 & $6.33 \times 10^{-6} \pm 1.36 \times 10^{-13}$ & $1.56 \pm 0.01$ & $760 \pm 0$ \\
& 10 & - & 32 & 1 & 320 & $1.33 \times 10^{-7} \pm 3.65 \times 10^{-14}$ & $3.82 \pm 0.02$ & $764 \pm 0$ \\
\midrule
\multirow{3}{*}{\parbox{2.3cm}{CP-PINN \\ (80\,K epochs)}} & 2 & $32^2$ & - & 4 & 216 & $1.45 \pm 0.21$ & $102.1 \pm 12.3$ & $882 \pm 0$ \\
& 2 & $1\text{,}024^2$ & - & 4 & 216 & $0.63 \pm 0.13$ & $315.5 \pm 10.7$ & $1\text{,}014 \pm 0$ \\
& 2 & $2\text{,}048^2$ & - & 4 & 296 & $0.60 \pm 0.26$ & $1\text{,}378 \pm 30.7$ & $1\text{,}014 \pm 0$ \\
\midrule
\multirow{3}{*}{\parbox{2.3cm}{KAN \\ (50 steps)}} & 2 & $32^2$ & 5 & - & 348 & $1.00 \pm 1.96 \times 10^{-5}$ & $38.9 \pm 3.05$ & $866 \pm 0$ \\
& 2 & $128^2$ & 10 & - & $5\text{,}498$ & $0.89 \pm 0.10$ & $191 \pm 4.81$ & $3\text{,}368 \pm 0$ \\
& 2 & $128^2$ & 20 & - & $28\text{,}578$ & $0.53 \pm 0.03$ & $369.6 \pm 2.06$ & $11\text{,}668 \pm 0$ \\
\bottomrule
\end{tabular}
} 
\label{tab:case2_domain_stats}
\end{table}

\begin{figure}[!hbt]
\centering
\includegraphics[width=1.0\linewidth]{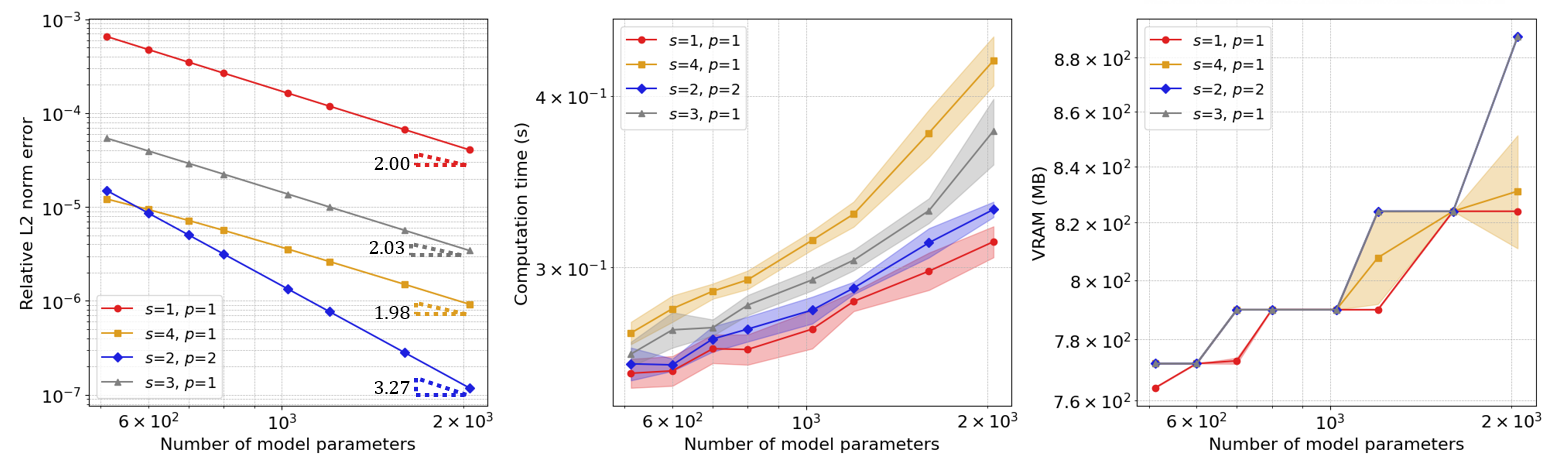}
\caption{Solving the 2D Poisson Eq. using INN-TD with different number of model parameters and hyperparameters $s$ and $p$. $a$=20 is used for all cases. The statistics are obtained by repeating each case for 10 times and the shaded area represents 1 standard deviation (a) convergence of INN-TD using different $s$ and $p$ (b) statistics of total computational time (c) statistics of GPU VRAM usage}
\label{poisson2dmultiply}
\end{figure}

Here we investigate the convergence of INN-TD for the 2D case with domain size $\Omega \in [0,12]^2$. The details are shown in Fig. \ref{poisson2dmultiply}. INN-TD shows a convergence property where a larger number of model parameters leads to better accuracy. Moreover, the convergence rate can be controlled by adopting different combinations of hyperparameters $s$ and $p$. 

\subsubsection{Solving the Helmholtz equation}
In this example, INN-TD is used to solve the Helmholtz equation as shown below. In \cite{wang2021understanding}, it is shown that the vanilla PINN 
exhibits failure modes when solving this equation. Here we use INN-TD to solve the exact same problem.
\begin{equation}\left(\frac{\partial^2}{\partial x_1^2}+\frac{\partial^2}{\partial x_2^2}\right)u+u(x,y)=q(x,y)\end{equation}

where the forcing function is defined as:
\begin{equation}q(x,y)=-(a_1\pi)^2\mathrm{sin}(a_1\pi x)\mathrm{sin}(a_2\pi y)-(a_2\pi)^2\mathrm{sin}(a_1\pi x)\mathrm{sin}(a_2\pi y)+k^2\mathrm{sin}(a_1\pi x)\mathrm{sin}(a_2\pi y)\end{equation}

The solution domain is defined as: $\Omega\in[-1,1]\times[-1,1]$. The analytical solution to this problem is:
\begin{equation}u(x_1,x_2)=\sin(a_1\pi x)\sin(a_2\pi y)\end{equation}

\begin{figure}[!hbt]
\centering
\includegraphics[width=0.85\linewidth]{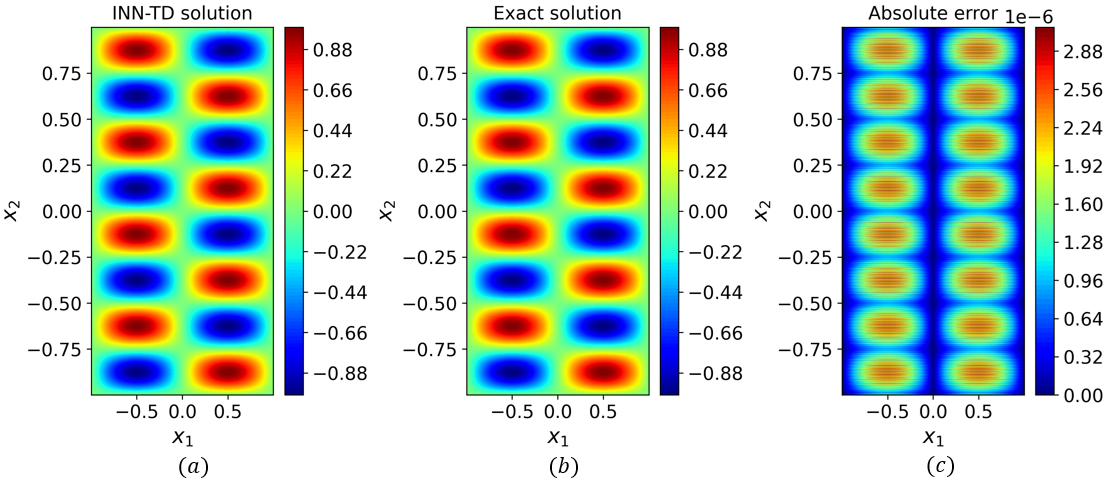}
\caption{Solving the 2D Helmholtz Eq. using INN-TD ($s=p=1, a=20$) with 2 modes and 250 elements along each direction. The total solution time is 0.48 sec (a) INN-TD solution (b) exact solution (c) point-wise absolute error}
\label{heml1}
\end{figure}

\begin{figure}[!hbt]
\centering
\includegraphics[width=0.95\linewidth]{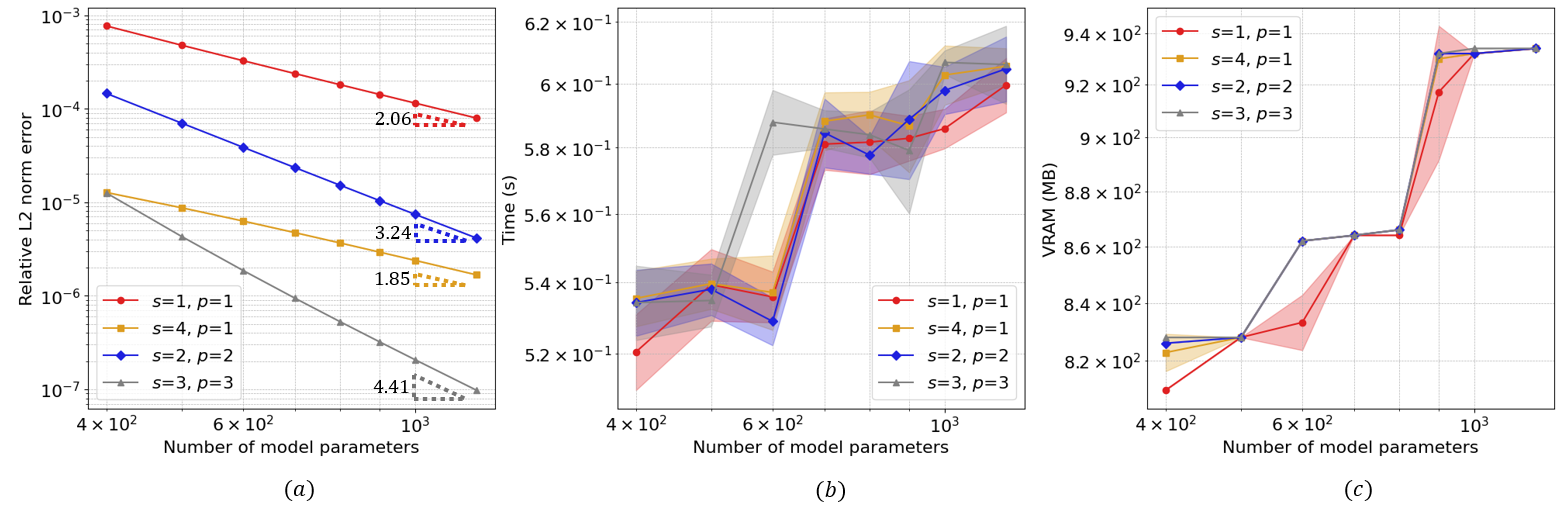}
\caption{Solving the 2D Helmholtz Eq. using INN-TD with different number of model parameters and hyperparameters $s$ and $p$. $a=20$ is used for all cases. The statistics are obtained by repeating each case for 10 times and the shaded area represents 1 standard deviation (a) convergence of INN-TD using different $s$ and $p$ (b) statistics of total computational time (c) statistics of GPU VRAM usage}
\label{heml2}
\end{figure}

 The comparison of INN-TD solution and the analytical solution is shown in Fig. \ref{heml1}. It can be seen that INN-TD achieves high accuracy without any failure modes. Furthermore, we study the convergence of INN-TD with different hyperparameters. The result is shown in Fig. \ref{heml2}. As expected, INN-TD shows the convergence property where a larger number of model parameters leads to higher accuracy. Higher $s$ and $p$ can also decrease the error.

\subsubsection{Space-time PDE example}
In this example, we show the convergence property of INN-TD data-free solver for a space-time problem. The governing PDE is shown below:
\begin{equation}
    \dot{u}(\boldsymbol{x},t)+k\Delta u(\boldsymbol{x},t)=b(\boldsymbol{x},t)\quad\mathrm{in} \quad \Omega_{\bm{x}}\otimes\Omega_t
\end{equation}
where $k$ is heat conductivity. $b(\boldsymbol{x},t)$ is defined as:
\begin{equation}
    b(\boldsymbol{x},t)=exp\left(-\frac{2\left(\left(x-x_0(t)\right)^2+(y-y_0(t))^2\right)}{r^2}\right)
\end{equation}
where $r$ is the standard deviation that characterizes the width of the heat source; $[x_0 (t),y_0 (t)]$ is the heat source center.
This equation models the heat transfer with a moving heat source.
The 4D space-time continuum is adopted as: $\Omega_{\bm{x}} = [-10,10]^3$ ; $\Omega_t =(0,0.1]$. 

To investigate the convergence of INN-TD for space-time problems, we use the following forcing function:
\begin{equation}
    \begin{aligned}b(\boldsymbol{x},t)&=2(1-2y^2)(1-e^{-15t})e^{-y^2-(100t-x-5)^2}\\&+2(1-2(100t-x-5)^2)(1-e^{-15t})e^{-y^2-(100t-x-5)^2}+(1\\&-e^{-15t})(200x+1000-20000t)e^{-y^2-(100t-x-5)^2}\\&-15e^{-15t}e^{-y^2-(100t-x-5)^2}\end{aligned}
\end{equation}

As a result, the exact solution can be written as:
\begin{equation}
    u^{exact}(\bm{x},t)=(1-e^{-15t})e^{-y^{2}-(x-100t-5)^{2}}
\end{equation}

\begin{figure}[!hbt]
\centering
\includegraphics[width=0.85\linewidth]{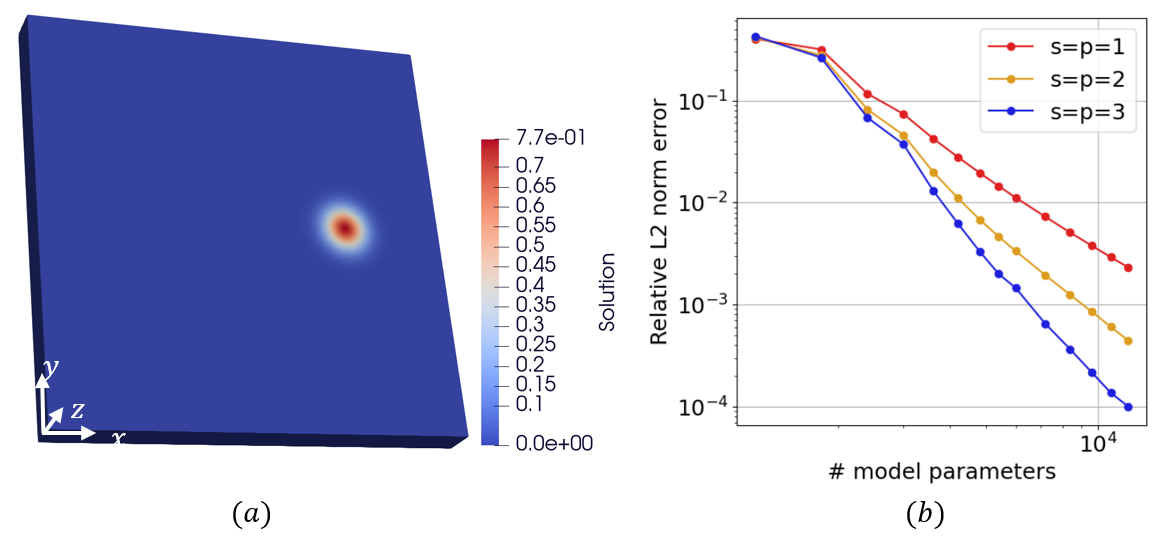}
\caption{Space-time PDE: (a) snapshot of the space-time solution (b) convergence for the 4D space-time problem: relationship between number of model parameters and relative L2 norm error}
\label{sp}
\end{figure}

A sample snapshot of the solution is shown in Fig. \ref{sp} (a). 

Since it's straightforward for numerical integration of C-HiDeNN interpolation function, here we define the relative L2 norm error in the integral form:
\begin{equation}
    \frac{\left(\int\left(u^{{exact}}(\bm{x},t)-\mathcal{J}u(\bm{x},t)\right)^2\mathrm{d}\bm{x}dt\right)^{\frac{1}{2}}}{\left(\int\left(u^{{exact}}(\bm{x})\right)^2\mathrm{d}\bm{x}dt\right)^{\frac{1}{2}}}
\end{equation}

We investigated the convergence properties of INN-TD using different hyperparameters $s$ and $p$ in C-HiDeNN interpolation. As shown in Fig. \ref{sp} (b), the error of INN-TD decreases as more model parameters are utilized. This guaranteed convergence distinguishes INN-TD from other black-box deep learning models, where an increase in model parameters does not necessarily result in reduced error. Additionally, the convergence rate can be controlled by adjusting $s$ and $p$, providing flexibility in designing the INN-TD according to the accuracy requirement.

\subsubsection{Space-parameter-time (S-P-T) PDE example}
\label{sec:spt}
To solve the 6D S-P-T problem, we used 101 grid points to discretize each input dimension of $(x, y, z, k, P, t)$. 100 modes were used for INN-TD approximation. Therefore, the INN-TD interpolation function can be written as:
\begin{equation}
    \mathcal{J}u(\bm{x})=\sum_{m=1}^{100}\widetilde{\boldsymbol{N}}_{x}(x)\boldsymbol{u}_{x}^{(m)}\cdot\widetilde{\boldsymbol{N}}_y(y)\boldsymbol{u}_{y}^{(m)}\cdot \widetilde{\boldsymbol{N}}_z(z)\boldsymbol{u}_{z}^{(m)} \cdot \widetilde{\boldsymbol{N}}_k(k)\boldsymbol{u}_{k}^{(m)} \cdot \widetilde{\boldsymbol{N}}_P(P)\boldsymbol{u}_{P}^{(m)} \cdot\widetilde{\boldsymbol{N}}_t(t)\boldsymbol{u}_{t}^{(m)}
    \label{eqtd}
\end{equation}

Algorithm \ref{algo:solver} was used to solve the S-P-T problem, and it took in total 155 s to run the solver on a single NVIDIA RTX A6000 GPU.

Since the analytical solution is not available for this problem, we used the finite element method (FEM) to generate the simulation data and treated it as the ground truth. For ease of validation, the same spatial mesh size ($100\times100\times100$) and 100 time steps were adopted in FEM simulation. We repeated running FEM simulations with 10 randomly selected parametric input pairs $(k, P)$. The relative L2 norm error is computed as follows:

\begin{equation}
    \epsilon=\left[\sum_{i,j,a,b,c,d}[\mathcal{J}u(x_i,y_j,z_a,k_b,P_c,t_d)-u^{FEM}(x_i,y_j,z_a,t_d; k_b,P_c)^{2}\right]^{\frac{1}{2}}/\left[\sum_{i,j,a,b,c,d}u^{FEM}(x_i,y_j,z_a,t_d; k_b,P_c)^{2}\right]^{\frac{1}{2}}
\label{error}
\end{equation}

where $i, j, a, b, c, d$ refer to the data index of each input dimension.
\subsubsection{Inverse optimization problem}
\label{sec:inverse}
In this problem, we solved the S-P-T problem using INN-TD data-free solver. As a result, we obtained the S-P-T interpolation function as shown in Eq. \ref{eqtd}. 

To test the effectiveness of INN-TD for the inverse optimization problem, we randomly sampled 100 different parametric input pairs $(k_i, P_i), i=1,..,100$, and generated the corresponding ground truth through high-fidelity FEM simulation. Then we used Adam optimizer to optimize the loss function as shown in Eq. \ref{inverse_loss}. The learning rate was chosen as 0.1. To ensure the parameters $(k_i, P_i)$ are within the predefined range, we added a box constraint on the parameter such that:
$$k_{min}\leq k_{i}<k_{max}$$
$$P_{min}\leq P_{i}<P_{max}$$

For each optimization case, we used uniform distributions for the initial guess. The mean and standard deviation for the relative L2 norm error and parameter error were then computed across all cases.

\subsubsection{Vector solution field: elasticity problem}
\label{sec:elasticity}
In this example, INN-TD is used to solve the 3D elasticity problem in solid mechanics where the solution field is a vector field. The solution $\bm{u} = [u_1(x,y,z), u_2(x,y,z), u_3(x,y,z)]$ represents the displacement along the $x,y$ and $z$ directions. The governing equation is shown below:
\begin{equation}\mu u_{i,jj}+(\mu+\lambda)u_{j,ji}+b_i=0\end{equation}
where Einstein summation is used in the index notation, $i = 1,2,3$ and $j=1,2,3$; $\lambda, \mu$ are the Lamé constants. The solution domain is defined as $\Omega = [0,1]\times[0,1]\times[0,2]$ and homogeneous boundary condition is assumed. The forcing function $b_i$ is defined as:
\begin{equation}\begin{cases}b_{1}&=922765196.706382\cdot n_{2}^{2}x(x-1)\sin(n_{1}\pi y)\sin\left(\frac{n_{1}\pi z}{2}\right)-58745053.109746\cdot n_{2}\cos(n_{2}\pi x)\left(2y-1\right)\sin\left(\frac{n_{2}\pi z}{2}\right)\\&-58745056.109746\cdot n_{3}\cos(n_{3}\pi y)\sin(n_{3}\pi y)\left(z-1\right)-523575701.269786\cdot\sin(n_{1}\pi y)\sin\left(\frac{n_{1}\pi z}{2}\right)\\b_{2}&=922765186.706382\cdot n_{2}^{2}\sin(n_{2}\pi x)\mathrm{~y(y-1)}\sin\left(\frac{n_{2}\pi z}{2}\right)-587450563.109746\cdot n_{1}(2x-1)\cos(n_{1}\pi y)\sin\left(\frac{n_{1}\pi z}{2}\right)\\&-587450563.109746\cdot n_{3}\sin(n_{3}\pi x)\cos(n_{3}\pi y)\left(z-1\right)-523575701.269788\cdot\sin(n_{2}\pi x)\sin\left(\frac{n_{2}\pi z}{2}\right)\\b_{3}&=738212149.36510\cdot n_{3}^{2}\sin(n_{3}\pi x)\sin(n_{3}\pi y)z(z-2)-293725281.554873\cdot n_{2}\sin(n_{2}\pi x)\left(2y-1\right)\cos\left(\frac{n_{2}\pi z}{2}\right)\\&-2937252181554873\cdot n_{1}(2x-1)\sin(n_{1}\pi y)\cos\left(\frac{n_{1}\pi z}{2}\right)-261787850.634994\cdot\sin(n_{3}\pi x)\sin(n_{3}\pi y)\end{cases}\end{equation}

The analytical solution to the problem is: 
\begin{equation}\begin{cases}u_1(x,y,z)=0.000972354873786749\cdot(x^2-x)\sin(n_1\pi y)\sin\left(\frac{n_1\pi z}{2}\right)\\u_2(x,y,z)=0.000972354873786749\cdot(y^2-y)\sin(n_2\pi x)\sin\left(\frac{n_2\pi z}{2}\right)\\u_3(x,y,z)=0.000972354873786749\cdot\sin(n_3\pi x)\sin(n_3\pi y)\left(\frac{z^2}{2}-z\right)\end{cases}\end{equation}

\begin{figure}[!hbt]
\centering
\includegraphics[width=0.85\linewidth]{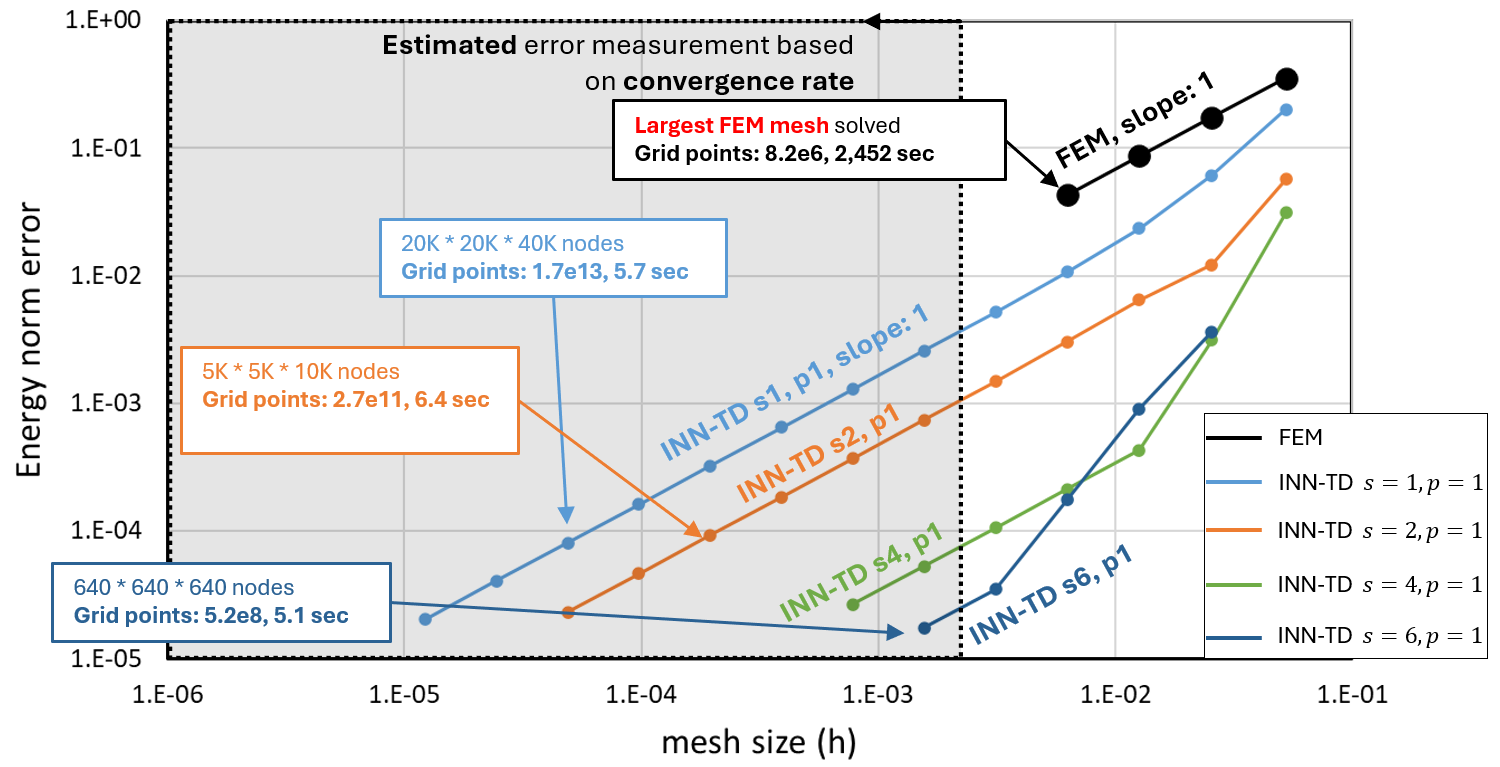}
\caption{Convergence for the elasticity problem: relationship between mesh size and energy norm error}
\label{elasticity}
\end{figure}

The convergence of INN-TD is shown in Fig. \ref{elasticity}.

\subsubsection{Operator solving}
INN-TD can also be extended to approximate PDE operators. As an example, we show that INN-TD can be leveraged to approximate the PDE operator for the following equation:
\begin{equation}\frac{\partial u}{\partial t}-\frac{\partial}{\partial x}k(x)\frac{\partial u}{\partial x}=f(x)\end{equation} with homogeneous initial and boundary conditions. Here we aim to approximate the PDE operator from the conductivity field $k(x)$ to the space-time PDE solution $u(x,t)$. 

\begin{figure}[!hbt]
\centering
\includegraphics[width=1.0\linewidth]{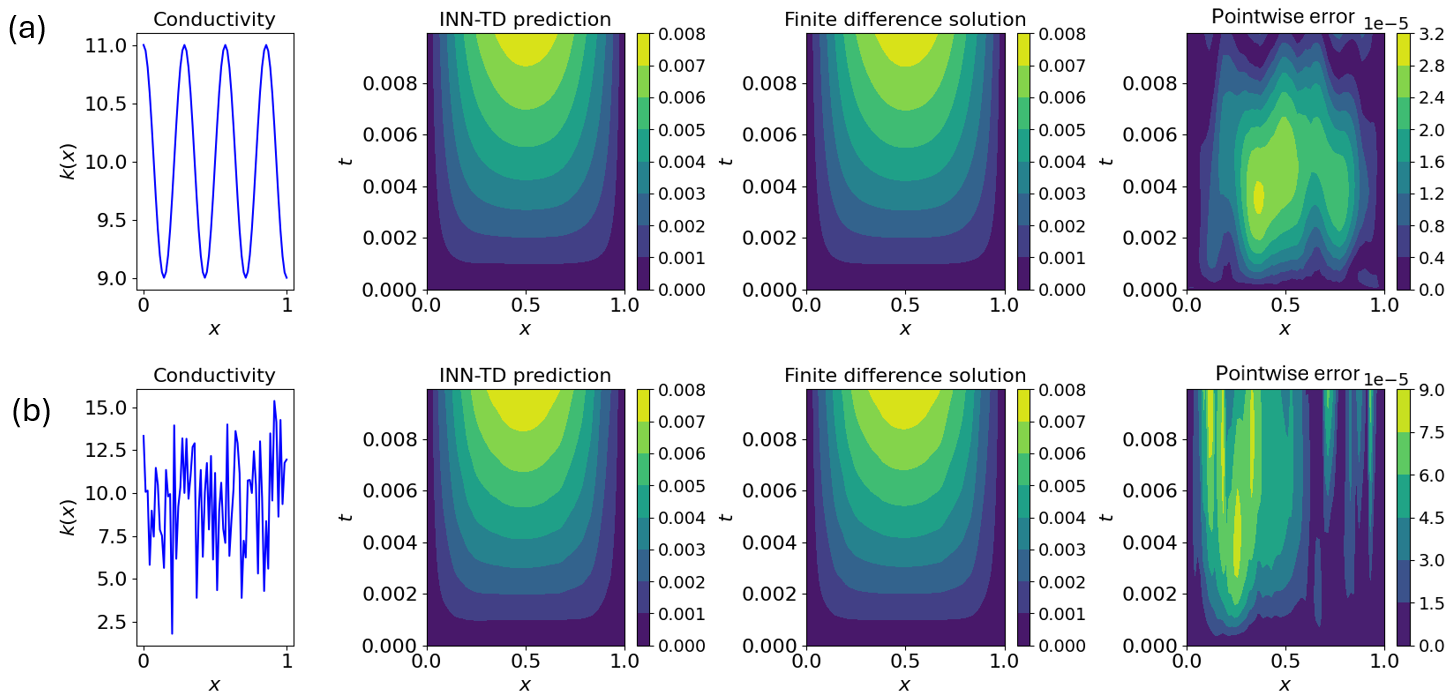}
\caption{Results for operator solving task when conductivity field $k(x)$ is a (a) sinusoidal function (b) random function.}
\label{fig:operator}
\end{figure}

To account for the arbitrariness of $k(x)$, we discretize $x$ and assume the nodal values of $k(x)$ are subjected to the following covariance function:
\begin{equation}C(x_i,x_j)=\sigma^2\exp\left(-\frac{\left(x_i-x_j\right)^2}{2l^2}\right)\end{equation}
where hyperparameters $\sigma$ and $l$ refer to standard deviation and length scale, respectively. The covariance matrix can be further decomposed using eigen-decomposition and arbitrary $k(x)$ is approximated using the following equation: 
\begin{equation}k(x,\bm{\zeta})=k_\mu+\sum_{I=1}^{n_x}\widetilde{N}_{I}(x)\sum_{J=1}^{n_e}\sqrt{\lambda_{J}}\phi_{IJ}\zeta_{J}\end{equation}
where $k_{\mu}$ is the mean value; $\lambda_J$ is the $J$-th eigenvalue; $\phi_{IJ}$ refers to the $I$-th component of $J$-th eigenvector; $\zeta_J$ is the $J$-th uncorrelated variable; $\widetilde{N}_I(x)$ is the $I$-th C-HiDeNN basis function. As a result, the INN-TD approximation of the PDE operator can be written as:
\begin{equation}\mathcal{J}u(x,\bm{\zeta},t)=\sum_{m=1}^Mu_x^{(m)}(x)\psi_1^{(m)}(\zeta_1)\psi_2^{(m)}(\zeta_2)\cdot...\cdot\psi_{n_e}^{(m)}(\zeta_{n_e})u_t^{(m)}(t)\end{equation}

where each univariate function is approximated using C-HiDeNN basis functions. The details of the model discretizations are shown in Table \ref{tab:operator}. INN-TD requires $96.74 \pm 4.65$ sec for the offline computation. Fig. \ref{fig:operator} shows that INN-TD gives accurate predictions as compared to the corresponding finite difference solutions for different realizations of the conductivity field.

\begin{table}[h]
\centering

\caption{Summary of the INN-TD model for approximating PDE operator}
\label{tab:operator}
\begin{tabular}{l|c|c|c}

\hline
\textbf{Discretization variable} & \textbf{Space (x)} & \textbf{Parameter for \( k \) (\(\zeta\))} & \textbf{Time (t)} \\
\hline
Domain & [0, 1] & [-5, 5] & [0, 0.01] \\
Number of independent variables & 1 & 71 & 1 \\
Number of elements & 71 & 101 & 151 \\
Number of eigenvalues & -- & 71 & -- \\
\hline
\end{tabular}
\end{table}

\end{document}